\documentstyle[12pt, epsfig]{article}

\advance \topmargin by -\headheight
\advance \topmargin by -\headsep
     
\evensidemargin \oddsidemargin
\begin{document}
\newcommand{\sect}[1]{\setcounter{equation}{0}\section{#1}}
\renewcommand{\theequation}{\thesection.\arabic{equation}}
\date{} 
\topmargin -.6in
\def\lab{\label}
\def\nonu{\nonumber}
\def\rf#1{(\ref{eq:#1})}
\def\lab#1{\label{eq:#1}} 
\def\br{\begin{eqnarray}}
\def\er{\end{eqnarray}}
\def\be{\begin{equation}}
\def\ee{\end{equation}}
\def\0{\nonumber}
\def\lb{\lbrack}
\def\rb{\rbrack}
\def\({\left(}
\def\){\right)}
\def\v{\vert}
\def\bv{\bigm\vert}
\def\lskip{\vskip\baselineskip\vskip-\parskip\noindent}
\relax
\newcommand{\nit}{\noindent}
\newcommand{\ct}[1]{\cite{#1}}
\newcommand{\bi}[1]{\bibitem{#1}}
\def\a{\alpha}
\def\b{\beta}
\def\ca{{\cal A}}
\def\cm{{\cal M}}
\def\cn{{\cal N}}
\def\cf{{\cal F}}
\def\d{\delta} 
\def\D{\Delta}
\def\eps{\epsilon}
\def\g{\gamma}
\def\G{\Gamma}
\def\grad{\nabla}
\def\h{ {1\over 2}  }
\def\hc{\hat{c}}
\def\hd{\hat{d}}
\def\hg{\hat{g}}
\def\hp{ {+{1\over 2}}  }
\def\hm{ {-{1\over 2}}  }
\def\k{\kappa}
\def\l{\lambda}
\def\L{\Lambda}
\def\lg{\langle}
\def\m{\mu}
\def\n{\nu}
\def\o{\over}
\def\om{\omega}
\def\O{\Omega}
\def\p{\phi}
\def\pa{\partial}
\def\pr{\prime}
\def\ra{\rightarrow}
\def\rh{\rho}
\def\rg{\rangle}
\def\s{\sigma}
\def\t{\tau}
\def\th{\theta}
\def\ti{\tilde}
\def\wti{\widetilde}
\def\inte{\int dx }
\def\xb{\bar{x}}
\def\yb{\bar{y}}
\def\tpsi{{\tilde \psi}}
\def\tchi{ {\tilde \chi}}
\def\btpsi{\bar {\tilde \psi}}
\def\btchi{\bar {\tilde \chi}}
\def\tr{\mathop{\rm tr}}
\def\Tr{\mathop{\rm Tr}}
\def\partder#1#2{{\partial #1\over\partial #2}}
\def\ds{{\cal D}_s}
\def\wtwo{{\wti W}_2}
\def\lie{{\cal G}}
\def\alie{{\widehat \lie}}
\def\dlie{{\cal G}^{\ast}}
\def\elie{{\widetilde \lie}}
\def\edlie{{\elie}^{\ast}}
\def\hlie{{\cal H}}
\def\wlie{{\widetilde \lie}}

\def\rlx{\relax\leavevmode}
\def\inbar{\vrule height1.5ex width.4pt depth0pt}
\def\IZ{\rlx\hbox{\sf Z\kern-.4em Z}}
\def\IR{\rlx\hbox{\rm I\kern-.18em R}}
\def\IC{\rlx\hbox{\,$\inbar\kern-.3em{\rm C}$}}
\def\one{\hbox{{1}\kern-.25em\hbox{l}}}

\def\PRL#1#2#3{{\sl Phys. Rev. Lett.} {\bf#1} (#2) #3}
\def\NPB#1#2#3{{\sl Nucl. Phys.} {\bf B#1} (#2) #3}
\def\NPBFS#1#2#3#4{{\sl Nucl. Phys.} {\bf B#2} [FS#1] (#3) #4}
\def\CMP#1#2#3{{\sl Commun. Math. Phys.} {\bf #1} (#2) #3}
\def\PRD#1#2#3{{\sl Phys. Rev.} {\bf D#1} (#2) #3}
\def\PLA#1#2#3{{\sl Phys. Lett.} {\bf #1A} (#2) #3}
\def\PLB#1#2#3{{\sl Phys. Lett.} {\bf #1B} (#2) #3}
\def\JMP#1#2#3{{\sl J. Math. Phys.} {\bf #1} (#2) #3}
\def\PTP#1#2#3{{\sl Prog. Theor. Phys.} {\bf #1} (#2) #3}
\def\SPTP#1#2#3{{\sl Suppl. Prog. Theor. Phys.} {\bf #1} (#2) #3}
\def\AoP#1#2#3{{\sl Ann. of Phys.} {\bf #1} (#2) #3}
\def\PNAS#1#2#3{{\sl Proc. Natl. Acad. Sci. USA} {\bf #1} (#2) #3}
\def\RMP#1#2#3{{\sl Rev. Mod. Phys.} {\bf #1} (#2) #3}
\def\PR#1#2#3{{\sl Phys. Reports} {\bf #1} (#2) #3}
\def\AoM#1#2#3{{\sl Ann. of Math.} {\bf #1} (#2) #3}
\def\UMN#1#2#3{{\sl Usp. Mat. Nauk} {\bf #1} (#2) #3}
\def\FAP#1#2#3{{\sl Funkt. Anal. Prilozheniya} {\bf #1} (#2) #3}
\def\FAaIA#1#2#3{{\sl Functional Analysis and Its Application} {\bf #1} (#2)
#3}
\def\BAMS#1#2#3{{\sl Bull. Am. Math. Soc.} {\bf #1} (#2) #3}
\def\TAMS#1#2#3{{\sl Trans. Am. Math. Soc.} {\bf #1} (#2) #3}
\def\InvM#1#2#3{{\sl Invent. Math.} {\bf #1} (#2) #3}
\def\LMP#1#2#3{{\sl Letters in Math. Phys.} {\bf #1} (#2) #3}
\def\IJMPA#1#2#3{{\sl Int. J. Mod. Phys.} {\bf A#1} (#2) #3}
\def\AdM#1#2#3{{\sl Advances in Math.} {\bf #1} (#2) #3}
\def\RMaP#1#2#3{{\sl Reports on Math. Phys.} {\bf #1} (#2) #3}
\def\IJM#1#2#3{{\sl Ill. J. Math.} {\bf #1} (#2) #3}
\def\APP#1#2#3{{\sl Acta Phys. Polon.} {\bf #1} (#2) #3}
\def\TMP#1#2#3{{\sl Theor. Mat. Phys.} {\bf #1} (#2) #3}
\def\JPA#1#2#3{{\sl J. Physics} {\bf A#1} (#2) #3}
\def\JSM#1#2#3{{\sl J. Soviet Math.} {\bf #1} (#2) #3}
\def\MPLA#1#2#3{{\sl Mod. Phys. Lett.} {\bf A#1} (#2) #3}
\def\JETP#1#2#3{{\sl Sov. Phys. JETP} {\bf #1} (#2) #3}
\def\JETPL#1#2#3{{\sl  Sov. Phys. JETP Lett.} {\bf #1} (#2) #3}
\def\PHSA#1#2#3{{\sl Physica} {\bf A#1} (#2) #3}
\newcommand\twomat[4]{\left(\begin{array}{cc}  
{#1} & {#2} \\ {#3} & {#4} \end{array} \right)}
\newcommand\twocol[2]{\left(\begin{array}{cc}  
{#1} \\ {#2} \end{array} \right)}
\newcommand\twovec[2]{\left(\begin{array}{cc}  
{#1} & {#2} \end{array} \right)}

\newcommand\threemat[9]{\left(\begin{array}{ccc}  
{#1} & {#2} & {#3}\\ {#4} & {#5} & {#6}\\ {#7} & {#8} & {#9} \end{array} \right)}
\newcommand\threecol[3]{\left(\begin{array}{ccc}  
{#1} \\ {#2} \\ {#3}\end{array} \right)}
\newcommand\threevec[3]{\left(\begin{array}{ccc}  
{#1} & {#2} & {#3}\end{array} \right)}

\newcommand\fourcol[4]{\left(\begin{array}{cccc}  
{#1} \\ {#2} \\ {#3} \\ {#4} \end{array} \right)}
\newcommand\fourvec[4]{\left(\begin{array}{cccc}  
{#1} & {#2} & {#3} & {#4} \end{array} \right)}


\begin{titlepage}
\vspace*{-2 cm}
\noindent
\begin{flushright}
\end{flushright}

\vskip 1 cm
\begin{center}
{\Large\bf Soliton Spectrum of Integrable Models with Local Symmetries  }\footnote{to appear in JHEP (2002)} \vglue 1  true cm
{ J.F. Gomes}${}^1$, E. P. Gueuvoghlanian${}^{2}$,
 { G.M. Sotkov}${}^1$ and { A.H. Zimerman}${}^1$\\

\vspace{1 cm}

${}^1${\footnotesize Instituto de F\'\i sica Te\'orica - IFT/UNESP\\
Rua Pamplona 145\\
01405-900, S\~ao Paulo - SP, Brazil}\\

${}^2${\footnotesize Departamento de Fisica de Particulas,\\
Faculdad de Fisica\\
Universidad de Santiago\\
E-15706 Santiago de Compustela, Spain}

\vspace{1 cm}

\end{center}

\normalsize
\vskip 0.2cm

\begin{center}
{\large {\bf ABSTRACT}}\\
\end{center}
\noindent
The soliton spectrum (massive and massless) of a family of integrable models with local $U(1)$ and $ U(1)\otimes
   U(1)$
 symmetries is studied.  These models represent relevant integrable deformations of 
  $SL(2,R) \otimes U(1)^{n-1}$ - WZW 
 and $SL(2,R) \otimes SL(2,R)\otimes U(1)^{n-2}$ - WZW models.  Their massless solitons 
 appear as specific topological
 solutions of the $U(1)$ (or $U(1)\otimes U(1)$) - CFTs.  The nonconformal analog of the 
 GKO-coset  formula is derived and
 used in the construction of the composite massive solitons of the ungauged integrable models.
  
\noindent

\vglue 1 true cm

\end{titlepage}

\sect{Introduction}
The $G$-WZW models and their gauged $G/H$-versions (for appropriated choice of  $G$ and $H$) are known to describe string
theories on curved backgrounds \cite{a}, \cite{horne}. 
 The simplest examples are the $SL(2,R)$ - WZW model, 
 representing string $AdS_3$ target
space time \cite{b} and $SL(2)/U(1)$ - WZW , giving rise to 2-d black hole geometry \cite{a}, \cite{horne}.  In this paper we consider specific
relevant perturbations ($\hat G, G_0, \mu_{ab}$) \cite{z1}\cite{z2} \cite{z3} of certain physicaly interesting (gauged)
 WZW models 
\br
{\cal L}^{pert}(\hat G, G_0) = {\cal L}_{G_0}^{WZW} (g_0) - {{k}\o {2\pi }}\mu_{ab} Tr (T_a g_0 T_b g_0^{-1})
\label{1.1}
\er
where $G_0 \subset \hat G$ is finite dimensional subgroup of the affine group $\hat G$ with generators $T_a, \;\; g_0 \in
G_0$ and $\mu_{ab}$ are real parameters.  They are expected to describe the {\it nonconformal} counterparts of the Maldacena
string/gauge theory correspondence \cite{mal} (i.e., deformations of $AdS_d/CFT_{d-1}$ \cite{ah}).  The first question to be addressed is about the
classical (and quantum) integrability of the models (\ref{1.1}), i.e., for a given affine group 
$\hat G$, how to chose $G_0
\subset \hat G$ and $\mu_{ab}$ such that (\ref{1.1}) is exactly integrable? Another important question concerns the nonperturbative
topologically stable solutions  of the model (\ref{1.1}) and their particle or strings interpretation.

The main purpose of the present  paper is the
description  of the semiclassical spectrum  of the finite energy topological
solutions (solitons and solitonic strings) of a class of integrable perturbations that preserve  (a) one local $U(1) (p=1)$
or (b) two  $U(1)\otimes U(1) (p=2)$ local symmetries.  The simplest representative  of type (a) integrable models (IMs)
 studied in this paper is given by the Lagrangian \cite{elek}
 \br
{\cal L}^{u}_{p=1} = {1\o 2}\eta_{ij} \pa \varphi_i \bar \pa \varphi_j + {{n}\o {2(n+1)}} \pa R_u \bar \pa R_u
+ \pa \chi_u \bar \pa \psi_u e^{\b (R_u -\varphi_1)} - V_u
\label{1.2}
\er
with potential 
\br
V_u = {{m^2}\o {\b^2}} \( \sum_{i=1}^{n-1} e^{\b (\varphi_{i-1} + \varphi_{i+1}- 2 \varphi_i)} + 
e^{\b (\varphi_{1} + \varphi_{n-1})}\( 1+ \b^2 \psi_u \chi_u e^{\b (R_u -\varphi_1)}\) -n \)
\nonu
\er
where $\varphi_0 = \varphi_n =0, i,j=1,2,\cdots n-1, \b^2 =-{{2\pi}\o {k}}$ and $ \eta_{ij} =2 \d_{ij}-\d_{i,j-1}-\d_{i,j+1}$ is the
$A_{n-1}$ Killing form.  The above ${\cal L}^{u}_{p=1} $ is indeed a special case of the 
${\cal L}^{pert}(\hat G, G_0)$
(\ref{1.1}) obtained by taking $\hat G = A_n^{(1)}, \quad G_0 = SL(2)\otimes U(1)^{n-1}, 
\quad \mu_{ab} = \mu_a \bar \mu_b$  and 
\br
\mu_aT_a = \eps_+ = m \(\sum_{i=2}^{n} E_{\a_i}^{(0)} + E_{-(\a_2+\cdots +\a_n)}^{(1)}\),\quad 
\bar \mu_b T_b = \eps_- = m \(\sum_{i=2}^{n} E_{-\a_i}^{(0)} + E_{(\a_2+\cdots +\a_n)}^{(-1)}\)
\label{1.22}
\er
The  fields $\varphi_i, R_u, \psi_u, \chi_u $ that appear in (\ref{1.2}) parametrize the $g_0 \in G_0$ group element  as follows
\br
g_0 = \exp \( \b \chi_u E_{-\a_1}^{(0)}\) \exp  \( \b \l_1 \cdot H^{(0)} R_u + \b \sum_{i=2}^n \varphi_{i-1}h_i^{(0)}\)
\exp \( \b \psi_u E_{\a_1}^{(0)}\)
\nonu
\er
An important property of these IMs is their invariance  under local $U(1)$ transformations ($\b = i\b_0$)
\br
R^{\pr}_u =R_u + \b_0 \(w(z) + \bar w(\bar z)\), \quad 
\psi_u^{\pr} = \psi_u e^{-i\b_0^2 w(z)}, \quad \chi_u^{\pr} = \chi_u e^{-i\b_0^2 \bar w(\bar z)}, \quad \varphi_j^{\pr} = 
\varphi_j
\label{1.3}
\er
where $w, \bar w$ are arbitrary chiral functions.   The IM (\ref{1.2})  represents $A_{n-1}$-abelian affine Toda theory
interacting with the thermal perturbation (with $\Phi_{ab}^{(j=1)}$ \cite{kz}) of the $SL(2,R)$-WZW model.  Two particular
cases should be mentioned:   $(1)$ $n=1$ (no $\varphi_i$ at all) and the IM (\ref{1.2}) just coincides with deformed
$SL(2,R)$-WZW model $(2)$ for $n=2$ (one $\varphi$ only) it gives rise to an integrable deformation of the
$SL(3,R)$-Bershadsky-Polyakov model \cite{ber},\cite{pol}, i.e. certain energy perturbation of the $W_3^{(2)}$-algebra minimal models
\cite{ber}.

An important feature of IM (\ref{1.2}) is that for imaginary coupling $\b =i \b_0$ (and $n\geq 2$) its potential $V_u$ has
n-distinct zeros and as a consequence the model (\ref{1.2}) admits $U(1)$-charged topological solitons. The main
characteristic of its soliton spectrum is the presence of massive ($j_{\varphi}, 0$) as well as massless  
($0, j_{w}$) solitons (and solitonic strings) interacting with each other ($j_{\varphi}, j_{w}$ are the topological charges
of fields $\varphi_l$ and $R_u$).  The massless sector is represented by $\lie_0^0 = U(1)$ chiral {\it conformal
field  theory} (CFT). 
In fact, together with the constant vacua solutions ($E=P=0, \;\; j_{\varphi} =j_{w} =0$)
\br
\varphi_l^{vac} = {{2\pi }\o {\b_0}}{{l}\o {n}}N, \quad \psi^{vac}\chi^{vac} =0, \quad R^{vac} = {{2\pi }\o {\b_0}}a_R
\label{1.4}
\er
($N\in Z$ and  $a_R$ is a real parameter),  the equations of motion of IM (\ref{1.2}) admits also
massless (1-D string-like) conformal solutions  with nonvanishing energy $E=\pm P \neq 0$, 
 and $j_{w} \neq 0$.  They have again the form (\ref{1.4}), but with $R^{vac} $ replaced by
$R^{CFT}$, i.e.,
\br
R^{CFT} = \b_0 \( w(z) +\bar w(\bar z) \)
 \label{1.6}
 \er
 The spectrum of the left-moving solitons of this chiral $U(1)$-CFT (with appropriate periodic b.c.'s for $w, \bar w$),
 derived in Sect. 3.4 has the form:
 \br
 E^{CFT}_{L-sol} &=& {{4n}\o {(n+1)\b_0^2}}\( s + {{j_{w}}\o n}\) ^2|a_0|, \nonu \\
 j_w &=& j_{\varphi} -2 j_q, \quad j_{\varphi}, j_q = 0, \pm 1, \cdots , \pm (n-1)\; {\rm mod }\;\;  n \nonu \\
 Q_0 &=&  {{4\pi n}\o {n+1}} \(s + {{j_w}\o n}\), \quad s=0, \pm 1, \cdots
 \label{1.7}
 \er
where $|a_0|$ is an arbitrary infrared mass scale and $Q_0$ denotes the left-$U(1)$ charge.  The conformal sector of the
theory provides each one of the $n$-vacua states ($E^v = 0$) with a tower of conformal states 
 ($j_{\varphi}, j_{w},s$) with $E^{CFT}\neq 0$.  The {\it massless} solitons are topologically stable solutions that interpolate 
between the vacua ($0,0,0$) and an arbitrary conformal state $(0,j_w,s)$.  Together with such massless solutions, it is
natural to expect the existence of two types of {\it massive} solitons:
\begin{enumerate}
\item interpolating between different vacua, called g-solitons: $(0,0,0) \rightarrow (j_{\varphi}, 0,0)$

\item interpolating between vacua and arbitrary exited conformal states, called u-solitons: 
$(0,0,0) \rightarrow (j_{\varphi}, j_w,s)$

\end{enumerate}
\newpage
as it is shown by the diagram

\begin{figure}[ht] \label{fig1}
    \centering
    \epsfig{file=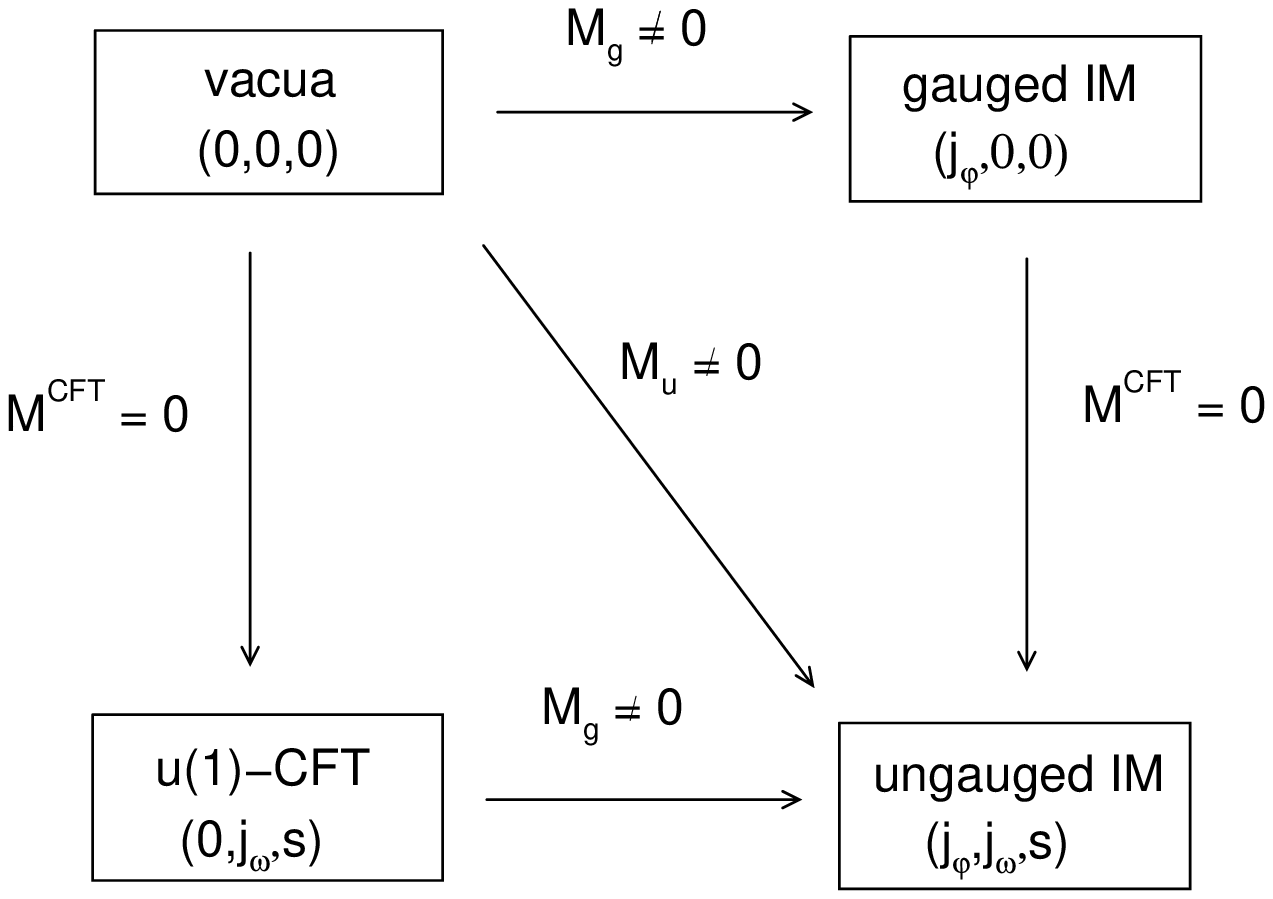,clip=,width=3.8in}
\end{figure}

In order to construct such soliton solutions and to derive their 
semi-classical spectrum, it is crucial to observe that the type $(1)$
g-solitons in fact coincide with the $U(1)$-charged topological solitons of the gauged dyonic IM \cite{elek},
\cite{tau} (with one global
$U(1)$ symmetry) obtained from (\ref{1.2}) by axial gauging of the local  $U(1)$:
\br
{\cal L}_g^{ax}(p=1) = {1\o 2} \sum_{i,j=1}^{n-1} \eta_{ij} \pa \varphi_i \bar \pa \varphi_j + {{\pa \chi_g \bar \pa \psi_g
e^{-\b \varphi_1}}\o {1+\b^2 {{n+1}\o {2n}} \psi_g \chi_ge^{-\b\varphi_1}}} - V_g, 
\label{1.9}
\er
\br
V_g = {{m^2}\o {\b^2}}\( \sum_{i=1}^{n-1} e^{-\b \eta_{ij} \varphi_j} + e^{\b (\varphi_1 + 
\varphi_{n-1}}( 1+\b^2 \psi_g \chi_ge^{-\b\varphi_1}) -n \)
\nonu 
\er
The semiclassical spectrum of the 1-solitons of this gauged IM has been derived in our recent paper 
\cite{elek}.

The type $(2)$ u-solitons turns out to be the conformal dressing of the g-solitons, i.e. by performing specific $U(1)$ gauge
transformations (\ref{1.3}) (with $w$ and $\bar w$ given by eqn. (\ref{3.40})) of the already known 
g-solitons \cite{elek}, \cite{tau}.  The
spectrum of these new u-solitons can be obtained by applying the following nonconformal version of the 
Goddard-Kent-Olive (GKO) coset construction \cite{god},
\br
T_{G_0}^{u} = T_{G_0/G_0^0}^{g} + T_{G_0^0}^{CFT}, \quad E^u = E^g + E^{CFT}
\label{1.10}
\er
establishing the relation  between the stress-tensors (and energies) of the ungauged IM (\ref{1.2}), the gauged
IMs (\ref{1.9}) and the $U(1)$-CFT.  The spectrum of the left-moving u-solitons ($w\neq 0, \bar w =0$), derived in Sect.
3.5, has the form,
\br
M_u^2 &=& M_g \( M_g + {{8nm_0}\o {(n+1)\b_0^2}}(s+{{j_w}\o{n}})^2\), \nonu \\
M_g &=& {{4 m n}\o {\b_0^2}}|\sin ({{4\pi j_{\varphi} -\b_0^2 j_{el}}\o {4n}})|, \quad 
Q_{el} = \b_0^2 j_{el}, \quad j_{el} = 0, \pm 1, \cdots  \nonu \\
Q_R^u &=& {{n+1}\o {2n}}\( Q_{el} + {{2\pi n}\o {n+1}} (s+{{j_{w}}\o n})\), \quad Q_{\theta}^u = 
{{\pi}\o 2}(s+{{j_{w}}\o n})
\nonu
\er
where $m_0 =|a_0|e^{b_L}$ and g-soliton velocity $v_g$ coincides with the left rapidity $b_L$, i.e., $v_g = b_L$.  The
stability of these strong coupling particles is ensured by their nontrivial topological charges:
 $j_{\varphi}, s, j_w$.

The paper is organized as follows. 
 In Sect. 2 we present a brief introduction to the path integral version of the
Hamiltonian reduction method.  The Lagrangians of the 
different IMs (with one local $U(1)$, two local $U(1)\otimes U(1)$
 and one local and one global $U(1)$'s ) considered in this paper are derived 
 together with the proof of their
 integrability.  Sect. 2.3 is devoted to the identification of the above IMs as 
 relevant integrable deformations of the
 conformal minimal models of certain extended conformal algebras $W_{n+1}^{(n)}$ 
 and $V_{n+1}^{(1,1)}$.  The discussion of
 Sect. 3 is concentrated on the symmetries (continuous $U(1)$ and discrete 
 $Z_n \otimes Z_2$), their currents and especially
 on the allowed b.c's for the fields of IMs (\ref{1.2}) and (\ref{1.9}).  The main result of Sect. 3.2 is the derivation of
 the {\it nonconformal} GKO coset construction.  In Sect. 3.4, we study the properties of the different $U(1)$-CFT solitons:
 left, right and left-right ones.  The spectrum of the composite u-solitons is obtained in Sect. 3.5.  The spectral flow of the
 $U(1)$-charges $Q_0, \bar Q_0$ induced from certain topological $\theta$-terms added to the original Lagrangian (\ref{1.2})
 is derived in Sect. 3.6.  Sect. 3.7 is devoted to the nontopological solitons of the deformed $SL(2,R)$-WZW model.  The
 constructions of different soliton solutions of the ungauged IM (\ref{2.13}) with $U(1)\otimes U(1)$ local symmetries as
 well as of the so called intermediate IM are presented in Sect. 4.  Sect. 5 contains our conclusions and few remarks
 concerning the use of the charge spectrum of our soliton and string solutions in the off-critical (i.e. nonconformal )
 $AdS_3/CFT_2$ correspondence.  The IMs with local $SL(2,R)\otimes U(1)$ symmetry and an example of nonrelativistic IM with
 local $U(1)$ symmetry are also discussed.

\sect{Integrable Perturbations of Gauged $A_n^{(1)}$-WZW Model }

\subsection{Effective Lagrangians from Hamiltonian reduction}

Different integrable perturbations ($\hat G, G_0, \mu_{ab}$) of the gauged WZW
model (\ref{1.1}) are known to be (at least classically) in one to one
correspondence to the admissible graded structures 
($\hat {\lie} , Q, \eps_{\pm}$) of the affine algrebra $\hat \lie $
\br
[Q, \lie_{k}] =k \lie_k, \quad [\lie_k, \lie_{l}] \subset \lie_{k+l}, \quad k,l \in
Z, \quad \lie_k \subset  \hat {\lie}, \quad [Q, \eps_{\pm }] =\pm  \eps_{\pm}
\nonu
\er
introduced by an appropriate grading operator \footnote{$\l_i$ are the fundamental weights  and  $\a_i \cdot H^{(a)}, 
E_{\a}^{(a)}$ the  generators of $\hat {\lie}$.  The operator  $d$ is the derivation operator.}
\br
Q=\tilde h d + \sum_{i=1}^{rank \hat {\lie}=n} s_i \l_i \cdot H^{(0)}, \quad [d, E_{\a}^{(a)}]= aE_{\a}^{(a)}, \;\; a, s_i, \tilde h \in Z
\label{2.1}
\er
such that the zero graded subalgebra $\lie_0$, i.e., $  [Q, \lie_0]=0$ is finite dimensional. 
The functional integral version of the Hamiltonian reduction method \cite{or}, \cite{annal1}, \cite{tau} consists in considering the 
two-loop (i.e., affine) $\hat {G}$-WZW model \cite{2loop} and imposing certain constraints on the $\hat {\lie}$-currents
$J_{\a}^{(a)} = Tr \( g^{-1} \pa g E_{\a}^{(a)} \)$ and 
$\bar J_{-\a}^{(a)} = Tr \( \bar \pa g g^{-1}E_{-\a}^{(a)} \)$ from the
 positive and negative grades 
$|q| \geq 1$ subalgebras \footnote{these are the simplest grade $|q| = 1$ IMs.  Imposing
 constraints on 
$J_{\a}^{(a)}, \bar J_{-\a}^{(a)}$ of grades $|q| \geq s$, one can construct in this way the so called higher grades  
IMs \cite{ferr}.}
${\cal H}_{\pm} = \oplus _{l\in Z_{\pm}} \hat {\lie}_{l}$.   More precisely, we
introduce the ${\cal H}_{\pm}$-invariant gauged ${\cal H}_{-}\backslash \hat {G}/ {\cal H}_{+}$-WZW model
\br
S(g,A_+,\bar{A}_-)&=&S_{WZW}^{\hat G}(g)
\nonumber
\\
&-&\frac{k}{2\pi}\int d^2x Tr\( A_-(\bar{\partial}gg^{-1}-\epsilon_{+})
+\bar{A}_+(g^{-1}\partial g-\epsilon_{-})+A_-g\bar{A}_+g^{-1}  \) \nonu \\
\label{2.2}
\er
where $\bar A_+(z, \bar z) \in {\cal H}_{+}, \; A_-(z, \bar z) \in {\cal H}_{-}, \; g(z, \bar z) \in \hat {G}$ and $\eps_{\pm}$
are specific constant linear combinations of grade $\pm 1$ generators of $\hat \lie$, say,
\br
\eps_+ = \mu_a T_a = \sum _{i} \mu_i E_{\a_i}^{(0)} + \mu_0 E_{-\s}^{(1)}, \quad 
\eps_- = \bar \mu_a T_a = \sum _{i} \bar \mu_i E_{-\a_i}^{(0)} + \bar \mu_0 E_{\s}^{(-1)}, \nonu
\er
where $\mu_{ab} = \mu_a \bar \mu_b$ and $\s $ is a fixed composite root, such that 
$[Q, E_{\mp \s}^{(\pm 1)}]=  \pm E_{\mp \s}^{(\pm 1)}$, $\a_i $ denote the simple roots of $\hat \lie$.  Then the effective Lagrangian  (\ref{1.1}) appears as a result of
performing the Gaussian integral on $A_-$ and $\bar A_+$ in the partition function
\br
{\cal Z}= \int Dg DA_- D\bar A_+ \exp (-S(g, A_-, \bar A_+)) \sim \int Dg_0 \exp (-S_{pert}(g_0))
\label{2.3}
\er
of the gauged two loop WZW model, where $g_0(z, \bar z) \in G_0 \subset \hat G$.  By construction for each admissible graded
structure (\ref{2.1}) of $\hat G$ and for any choice of $\eps_{\pm}$ the corresponding ${\cal L}_{pert}$ (\ref{1.1}) represents an
integrable model, as we shall demonstrate in the next Subsection 2.2.  Therefore the individual properties of such IMs are
determined by $G_0$ and $\eps_{\pm}$, containing all the information about the physical fields $g_0 = g_0(\varphi_l, \psi_a,
\chi_a, \cdots )$ and their  nonconformal interactions.  Depending on the grading operator $Q$ the zero grade subgroup $G_0
\subset \hat G$ can be {\it abelian } $G_0 = U(1)^l, \; 0 \leq l \leq n$, or {\it nonabelian}, say $ G_0 = SL(2)^p \otimes
U(1)^{n-p}$.  For example for $\hat G = A_n^{(1)}$, taking $Q$ in the form (principal gradation)  $Q=(n+1)d + \sum_{i=1}^n
\l_i \cdot H^{(0)}$ one conclude that $G_0 = U(1)^n$.  We next choose the most general $\eps_{\pm}$, such that $[Q,
\eps_{\pm}] =\pm \eps_{\pm}$, 
\br
\eps_{\pm} = \sum_{i=1}^n \mu_i^{\pm} E_{\pm \a_i}^{(0)} + \mu_0^{\pm} E_{\mp (\a_1+ \cdots +\a_n)}^{(\pm 1)}
\label{2.5}
\er
and observe that the corresponding ${\cal L}_{pert}$ (\ref{1.1}) takes the form of the well known 
abelian affine Toda model
\cite{lez-sav}.  It represents an integrable perturbation of the $W_{n+1}$-minimal models 
\cite{vega} which describes
marginal ($\mu_i^{\pm}$) and relevant ($\mu_0^{\pm}$) perturbations of a string on flat 
background with certain tachyonic
potential.  An interesting nonflat strings backgrounds are provided by the non-abelian 
(NA) affine Toda models
\cite{lez-sav},
i.e., when $G_0 \subset \hat G$ is non-abelian.  The simplest example of such model is 
defined by the following algebraic
data:
\br
\hat G = A_n^{(1)}, \quad G_0 &=& SL(2)\otimes U(1)^{n-1}, \quad Q= nd + \sum_{i=2}^{n} \l_i \cdot H^{(0)}, \nonu \\
\eps_{\pm} &=& m \sum_{i=2}^n E_{\pm \a_i}^{(0)} + E_{\mp (\a_2 +\cdots +\a_n)}^{(\pm )},  
\label{2.6}
\er 
and physical fields parametrizing the zero grade group element
\br
g_0 =e^{\b \chi_u E_{-\a_1}^{(0)}} e^{\b \l_1\cdot H R_u + \b \sum_{i=1}^{n-1}\varphi_i h_{i+1}}e^{\b \psi_u E_{\a_1}^{(0)}}
\nonu
\er
Its Lagrangian  derived from eqns. (\ref{2.2}) and (\ref{2.3}) has the form (\ref{1.2}) and represents an integrable perturbation of
$G_0$-WZW model.   The main new feature of this $A_n^{(1)}(p=1)$ affine NA-Toda model in comparison with the abelian one
defined by eqn.  (\ref{2.5}),
is the existence of nontrivial invariant subalgebra $\lie_0^0 \subset \lie_0$, such that $[\lie_0^0, \eps_{\pm}]=0$, i.e. 
$\lie_0^0 = U(1) = \{ \l_1\cdot H^{(0)}\}$.  It contains all the information about the continuous symmetries of the model. 
For the abelian affine Toda we have $\lie_0^0 = \emptyset $, i.e., no continuous symmetries.  In the NA-Toda case, (i.e.,
eqn. (\ref{1.1}) and (\ref{2.6}) leading to (\ref{1.2})) one can easely verify that the Lagrangian (\ref{1.1}) (and therefore 
(\ref{1.2})) is invariant under {\it chiral} $\lie_0^0 = U(1)$ transformations;
\br
g_0^{\pr} = e^{\b \bar w(\bar z)\l_1 \cdot H} g_0 e^{\b  w(z)\l_1 \cdot H}
\label{2.7}
\er
which is the compact form  of the field transformation (\ref{1.3}).  

Similarly to the {\it conformal} (i.e. unperturbed) WZW
models one can further gauge fix the above chiral $\lie_0^0$-symmetry.  
The standard procedure of gauge fixing \cite{gaw}
consists in considering the following ``improved'' action by the addition
 of auxiliary $\lie_0^0$-fields $A_0(z, \bar z), 
\bar A_0(z, \bar z)$, playing the role of Lagrange multipliers:
\br
S(g_0,A_0,\bar{A}_0)&=&S_{u}(g_0)
\nonumber
\\
&-&\frac{k}{2\pi}\int d^2x Tr\(\pm  A_0\bar{\partial}g_0g_0^{-1}
+\bar{A}_0 g_0^{-1}\partial g_0 \pm  A_0g_0\bar{A}_0g_0^{-1} + A_0 \bar A_0 \) \nonu \\
\label{2.8}
\er
where the signs $\pm $ takes place for axial/vector gaugings of the $U(1)$, respectively.  The new $A_0, \bar A_0$-dependent
terms added to the action $S_u(g_0)$ are responsible for imposing the additional constraints $J_{\l_1\cdot H} = Tr
\( \l_1\cdot H g_0^{-1}\partial g_0\) = \bar J_{\l_1\cdot H} = Tr
\( \l_1\cdot H \bar \partial g_0 g_0^{-1}\) =0$.  They promote the chiral $U(1)$ symmetry (\ref{2.7}) to the following local
$U(1)$ symmetry $\a_0 = \a_0(z, \bar z)\in U(1)$
\br
g_0^{\pr} = \a_0 g_0 \a_0^{\pr}, \quad \bar A_0^{\pr} = \bar A_0 -\bar \pa \a_0^{\pr}(\a_0^{\pr})^{-1}, \quad
 A_0^{\pr} =  A_0 - \a_0^{-1}\pa \a_0
\label{2.9}
\er
where $\a_0^{\pr} =\a_0$ in the case of axial gauging and $\a_0^{\pr} =\a_0^{-1}$ for the vector gauging.  Again, one can
take the Gaussian integral on $A_0, \bar A_0$ as in eqn. (\ref{2.3}) and the result is the effective Lagrangian (\ref{1.9})
of the axial singular affine NA-Toda denoted as dyonic $A_n^{(1)}(p=1)$ IMs ($p=1$ is the number of the gauged fixed
$U(1)$'s).

According to refs. \cite{elek}, \cite{annal2} the CPT invariant vector gauged Lagrangian  has the following form: 
\br
{\cal {L}}_g^{v}(p=1) &=& {1\o 2} \sum_{l=1}^{n-1} \(\pa \phi_l \bar \pa (\phi_l + \phi_{l+1} \cdots + \phi_{n-1}) + 
\bar \pa \phi_l \pa (\phi_l + \phi_{l+1} \cdots + \phi_{n-1})\) \nonu \\
&-&{1\o 2}{{\pa A \bar \pa B + \bar \pa A  \pa B}\o {1-\b^2 AB}} - V_{v}, \nonu 
\er
and 
\br
V_{v} &=& {{m^2}\o {\b^2}} \( A e^{\b (2\phi_1+\phi_2 +\cdots + \phi_{n-1})} + B e^{-\b (2 \phi_{n-1} + \phi_{n-2} +\cdots
\phi_1)} \) \nonu \\
&+& {{m^2}\o {\b^2}} \( \sum_{k=1}^{n-1} e^{\b (\phi_{k+1} - \phi_k)} -n \)
\label{2.10}
\er
The group element $g_0^v \in SL(2)\otimes U(1)^{n-1}/U(1)$ in  the vector case is parametrized by the fields $A,B$ and $\phi_l$ as follows
\br
 g_0^v = \twomat{d_2}{0}{0}{d_{n-1}},
  \quad d_{n-1} = diag (e^{\b \phi_1}, \cdots ,e^{\b \phi_{n-1}}), \quad
  d_2 = \twomat{A}{u}{{{AB-1}\o {ue^{\b (\phi_1+ \cdots +\phi_{n-1})}}}}{{{Be^{-\b (\phi_1+ \cdots +\phi_{n-1})}}}}
 \nonu
 \er
The nonlocal field $u$ (i.e., the vector analog of $R_g$) is defined by the following first order equations
\cite{elek}:
\br
\pa ln \( ue^{\b (\phi_1+ \cdots +\phi_{n-1})} \) = -\b^2 {{(A \pa B)}\o {1-\b^2 AB}}, \quad 
\bar \pa ln  u  = -\b^2 {{(B\bar \pa A)}\o {1-\b^2 AB}}.
\nonu
 \er
The main difference between ``ungauged'' IM ${\cal {L}}^u(p=1)$ (\ref{1.2}) and its gauged, axial (\ref{1.9})
and vector (\ref{2.10}) versions,is that the latters are invariant only under {\it global} $U(1)$ transformations, say ($\b = i \b_0$)
\br
A^{\pr} = e^{ia\b_0^2}A, \quad B^{\pr} = e^{-ia\b_0^2}B, \quad \phi^{\pr}_{l} = \phi_{l}- {{a \o n}}\b_0
\label{2.11}
\er
They contain one less field (no $R_u$ or $u$) and represent more complicated target space 
metric $g_{AB}\sim {{1\o {1-\b ^2 AB}}}$, than 
 the one of the ungauged model (\ref{1.2}).   The reason we
are considering them once more in the present paper is due to the crucial role they are going 
to play (see Sect. 3.3) in the
description of different 1-soliton solutions of the ungauged IM (\ref{1.2}).

Together with the IMs (\ref{1.2}) with one {\it local } $U(1)$ symmetry we shall study 
the soliton spectrum of another class
of {\it multicharged} $A_n^{(1)}(p=2)$ IMs \cite{new}, \cite{wigner} with two local $U(1)\otimes U(1)$ 
symmetries and also IMs with one local
and one global $U(1)$ described by Lagrangian (\ref{2.16}) below.  The starting point
 in the descrition of their effective Lagrangians
is the following specific {\it graded structure} of $\hat {G} = A_n^{(1)}$:
\br
Q&=&(n-1)d +\sum_{i=2}^{n-1} \l_i\cdot H^{(0)}, \quad G_0 = SL(2)\otimes SL(2)\otimes  U(1)^{n-2}, \quad \lie_0^0 =U(1)\otimes
U(1), \nonu \\
\eps_{\pm} &=& m \( \sum_{i=2}^{n-1}E_{\pm \a_i}^{(0)} + E_{\mp (\a_2 + \cdots + \a_{n-1})}^{(\pm 1)}\), \nonu \\
g_0 &=&e^{\b \sum_{a=1,n}\chi_a^u E_{-\a_a}^{(0)}} e^{\b \sum_{a=1,n}\l_a\cdot H R_a^u + 
\b \sum_{i=1}^{n-2}\varphi_i h_{i+1}}e^{\b \sum_{a=1,n}\psi_a^u E_{\a_a}^{(0)}}
\label {2.12}
\er
This algebraic data was used in deriving   the Lagrangian (\ref{2.2}) for the
 gauged fixed ${\cal H}_{\pm}$ two-loop WZW model.
Again, by integrating the auxiliary fields $A_+, \bar A_-$,  one gets (\ref{1.1}). 
 More explicitly  the
ungauged multicharged $A_n^{(1)}(p=2)$ IM is given by \cite{new}
\br
{\cal L}^u_{p=2} &=& {1\o 2}\sum_{i=1}^{n-2} \eta_{ij} \pa \varphi_i \bar \pa \varphi_j + \pa \chi_1^u \bar \pa \psi_1^u
e^{\b (R_1^u - \varphi_1)} 
+ \pa \chi_n^u \bar \pa \psi_n^u e^{\b (R_n^u - \varphi_{n-2})} \nonu \\
&+& {{1\o {2(n+1)}}}\( n\pa R_1^u \bar \pa R_1^u  
+n\pa R_n^u \bar \pa R_n^u + \pa R_1^u \bar \pa R_n^u +\pa R_n^u \bar \pa R_1^u \) -V_u^{p=2}
\label{2.13}
\er
with potential
\br
V_u^{p=2} &=& {{m^2}\o {\b^2}} \( \sum_{i=1}^{n-2} e^{-\b \eta_{ij} \varphi_j } +  e^{\b (\varphi_1+\varphi_{n-2})} (1+ \b^2
\psi_1^u \chi_1^u e^{\b (R_1^u - \varphi_1)})\right. \nonu\\
 &\times& \left. (1+ \b^2 \psi_n^u \chi_n^u e^{\b (R_n^u - \varphi_{n-2})}) -n+1\),  \label{2.132}
\er
where $\varphi_0 = \varphi_{n-1}=0$.  For the specific choice of  constant grade $\pm 1$ elements $\eps_{\pm}$ (\ref{2.12}),
the form of the invariant subalgebra $\lie_0^0 = \{\l_1\cdot H, \l_n\cdot H \}$ is an indication that the IM (\ref{2.13}) is 
 invariant under the following 
chiral $U(1)\otimes U(1)$  transformations
\br
g_0^{\pr}  =e^{\b \sum_{a} \l_a \cdot H \bar w_a(\bar z)} g_0 e^{\b \sum_{a} \l_a \cdot H  w_a(z)}
\label{2.14}
\er
with $\bar \pa w_a = \pa \bar w_a =0$.  The field transformation encoded in eq. (\ref{2.14}) have the form
\br
(R_a^{ u})^{\pr} = R_a^{ u} + \b_0 (w_a + \bar w_a), \quad \varphi^{\pr}_l = \varphi_l, \quad 
(\psi_a^{ u})^{\pr} = e^{-i\b_0^2 w_a }\psi_a^{ u}, \quad (\chi_a^{u})^{\pr}= e^{-i\b_0^2 \bar w_a }\chi_a^{ u}
\label{2.15}
\er

By axial or vector gauge fixing one of the local symmetries (say, the one generated by $(\l_1+\l_n)\cdot H$) one can derive
an interesting ``intermediate'' IM with one local and one global $U(1)$ symmetries.  We take eqn. (\ref{2.8})
with
$g_0 \in G_0 = SL(2)\otimes SL(2)\otimes U(1)^{n-2}$ and $A_0 = a_0(z, \bar z) (\l_1+\l_n)\cdot H, \; 
\bar A_0 = \bar a_0(z, \bar z) (\l_1+\l_n)\cdot H$, ($a_0, \bar a_0$ are arbitrary functions) and by performing the Gaussian
integration over  $a_0, \bar a_0$ we obtain the effective Lagrangian for the intermediate axial IM \cite{new}
\br
{\cal L}^{interm}_{p=2} &=& {{n-1}\o {n+1}}\pa \bar R \bar \pa \bar R +
 {1\o 2}\sum_{i=1}^{n-2} \eta_{ij}\pa \varphi_i \bar \pa \varphi_j  \nonu \\
& +& 
 {{1\o {\Delta_0}}} \( (1+{{\b^2}\o {4}} \bar \psi_n \bar \chi_n e^{-\b (\varphi_{n-2} +\bar R)}) 
 \bar \pa \bar \psi_1 \pa \bar
 \chi_1 e^{\b (\bar R - \varphi_1)} \right. \nonu \\
 &+& \left.   
(1+{{\b^2}\o {4}} \bar \psi_1 \bar \chi_1 e^{-\b (\varphi_{1} +\bar R)}) \bar \pa \bar \psi_n \pa \bar
 \chi_n e^{-\b (\bar R + \varphi_{n-2})}- {{\b^2}\o 4} (\bar \psi_n \bar \chi_1 \bar \pa \bar \psi_1 \pa \bar \chi_n 
 \right. \nonu \\
 &+& \left. 
 \bar \psi_1 \bar \chi_n \bar \pa \bar \psi_n \pa \bar \chi_1)e^{-\b (\varphi_1 + \varphi_{n-2})}\) - V_{interm}
 \label{2.16}
 \er
 with potential
 \br
 V_{interm} = {{m^2}\o {\b^2}} \( \sum_{i=1}^{n-2} e^{-\b \eta_{ij} \varphi_j } + e^{\b (\varphi_1 + \varphi_{n-2})} (1+
 \b^2 \bar \psi_1 \bar \chi_1e^{\b (\bar R-\varphi_1)})(1+\b^2 \bar \psi_n \bar \chi_n e^{-\b (\bar R+\varphi_{n-2})}) - n+1\)
\nonu
\er
and the denominator $\Delta_0$ is quadratic in $\bar \psi_a, \bar \chi_a$:
\br
\Delta_0= 1+ {{\b_0^2}\o 4} \( \bar \psi_1\bar \chi_1 e^{\b (\bar R -\varphi_1)} + 
\bar \psi_n\bar \chi_n e^{-\b (\bar R +\varphi_{n-2})}\)
\nonu
\er
The relations between the fields of the ungauged (\ref{2.13}) and the above axial gauged IM (\ref{2.16}) has the form
\br
\bar \psi_a = \psi^u_a e^{{{\b }\o 4}(R_1^u +R_n^u)}, \quad \bar \chi_a = \chi^u_a e^{{{\b }\o 4}(R_1^u +R_n^u)}, \quad
\bar R = {1\o 2}(R_1^u - R_n^u)
\label{2.17}
\er
as one can see by comparing the group elements $g_0 \in G_0 = SL(2)\otimes SL(2)\otimes  U(1)^{n-2}$ from eqn.
(\ref{2.12}) with $g_0^f \in G_0/ U(1)$ (for axial gauging), i.e.,
\br
g_0 = e^{{{\b }\o 2}(\l_n \cdot H R_1^u +\l_1 \cdot H R_n^u)} g_0^{f,interm}
e^{{{\b }\o 2}(\l_n \cdot H R_1^u +\l_1 \cdot H R_n^u)} 
\label{2.172}
\er
where $g_0^{f,interm}=e^{\b \sum_{a=1,n}\bar \chi_a E_{-\a_a}^{(0)}} 
e^{\b (\l_1 -\l_n)\cdot H \bar R + \b \sum_{i=1}^{n-2}\varphi_i h_{i+1}}e^{\b \sum_{a=1,n}\bar \psi_a E_{\a_a}^{(0)}}$.  Note that the $p=2$
Lagrangian (\ref{2.16}) is quite similar to (\ref{1.9}) of the $p=1$ gauged IM.  Both are invariant under one global $U(1)$
symmetry.  The denominators of both are quadratic in $\psi_a, \chi_a$, but (\ref{2.16}) involves an extra pair  $\psi^u_n,
\chi^u_n$ of charged fields and it is invariant under the following chiral $U(1)$ transformation:
\br
\bar \psi_1^{\pr } = e^{-i\b_0^2 w} \psi_1, \quad 
\bar \chi_1^{\pr } &=& e^{-i\b_0^2 \bar w} \chi_1, \quad
\bar \psi_n^{ \pr} = e^{i\b_0^2 w} \psi_n, \quad 
\bar \chi_n^{ \pr} = e^{i\b_0^2 \bar w} \chi_n, \nonu \\
\bar R^{\pr} &=& \bar R+ \b_0 (w + \bar w), \quad \varphi^{\pr}_l = \varphi_l, 
\label{2.18}
\er
The description of the 1-soliton spectrum of this IM (\ref{2.16}) is one of the main purposes of the present paper.  As we
shall show in the next Sect. 4.3, the structure of its soliton solutions is quite similar to the ones of $p=1$ ungauged IM
(\ref{1.2}).  Due to the common local $U(1)$ symmetry, they share the same $U(1)$ CFT, but their generic $M_u$-solitons are
conformal dressing of different ``gauged'' $ M_g$-solitons.  In the case of IM (\ref{2.16}) one need to know the
multicharged solitons of the corresponding completely gauged IM, i.e. the IM obtained by (axial) gauge fixing of both local
$U(1)\otimes U(1)$ symmetries of the ungauged IM (\ref{2.13}).  The procedure of gauge fixing is the same as in the case of intermediate
IM(\ref{2.16}), but now the auxiliary fields $A_0, \bar A_0\in \lie _0^0$ are different, namely, 
\br
A_0 = a_{01}\l_1\cdot H + a_{0n}\l_n\cdot H, \quad 
\bar A_0 = \bar a_{01}\l_1\cdot H + \bar a_{0n}\l_n\cdot H
\nonu 
\er
The result of the matrix Gaussian integration (over $a_{0a}, \bar a_{0a}, a=1,n$) is the following effective Lagrangian of the 
completely axialy gauged IM \cite{new}
\br
{\cal L}_{n}^{p=2} &=& {1\o 2} \sum_{i=1}^{n-2} 
\eta_{ij}\pa \varphi_i \bar \pa \varphi_j + {{1\o {\Delta}}} \( (1+ \b^2 {{n}\o {2(n-1)}}\psi_n \chi_n e^{-\b
\varphi_{n-2}})\bar \pa \psi_1 \pa \chi_1 e^{-\b \varphi_1} \right. \nonu \\
&+& \left. (1+ \b^2 {{n}\o {2(n-1)}}\psi_1 \chi_1 e^{-\b
\varphi_{1}})\bar \pa \psi_n \pa \chi_n e^{-\b \varphi_{n-2}}\right. \nonu \\
&+& \left. {{\b^2}\o {2(n-1)}} (\chi_1 \psi_n \bar \pa \psi_1 \pa \chi_n + 
\chi_n \psi_1 \bar \pa \psi_n \pa \chi_1)e^{-\b (\varphi_1 + \varphi_{n-2})} \)
 - V_n^{p=2}
 \label{2.19}
 \er
with potential
\br
V_n^{p=2} = {{\mu^2}\o {\b^2}}\( \sum_{i=1}^{n-2} e^{-\b \eta_{ij}\varphi_j} + e^{\b (\varphi_1 + \varphi_{n-2})}(1+ \b^2 \psi_n
\chi_n e^{-\b\varphi_{n-2}}) (1+\b^2 \psi_1\chi_1 e^{-\b\varphi_1}) -n+1 \)
\nonu
\er
where $\varphi_0 = \varphi_{n-1} =0$ and 
\br
 \Delta = 1+ {{\b^2n}\o {2(n-1)}}(\psi_1\chi_1 e^{-\b\varphi_1}+ 
\psi_n \chi_n e^{-\b\varphi_{n-2}}) + {{\b^4 (n+1)}\o {4(n-1)}}\psi_1 \chi_1 \psi_n \chi_n e^{-\b (\varphi_1 +
\varphi_{n-2})}.
\nonu
\er
The fields $\psi_a, \chi_a$ are related to the charged fields $ \psi_a^u ,\chi_a^u $ of the ungauged  
IM as follows:
\br
g_0 = e^{-{1\o 2} \l_a\cdot H R_a} g_0^{f, axial}  e^{-{1\o 2} \l_a\cdot H R_a}
\label{2.192}
\er
i.e., $\psi_a = \psi_a^u e^{{1\o 2}R_a}, \;\; \chi_a = \chi_a^u e^{{1\o 2}R_a}$.  The IM (\ref{2.19}) is invariant under global
$U(1)\otimes U(1)$ transformations only ($\eps_a$ = const),
\br
\psi_a^{\pr} = \psi_a e^{i\b^2\eps_a}, \quad  \chi_a^{\pr} = \chi_a^u e^{-i\b^2\eps_a}, \quad \varphi_l^{\pr} = \varphi_l
\label{2.20}
\er
Its charged $(Q_1, Q_n)$ 1-solitons $M_g(p=2)$ \cite{new} appears as the basic ingredient in the construction of the 1-solitons
of both the ungauged $p=2$ IM (\ref{2.13}) and the intermediate IM (\ref{2.16}).

\subsection{Zero curvarure representation and  classical integrability}

The proof of the (classical) integrability of the 2-d models (\ref{1.2}), (\ref{1.9}),(\ref{2.13}),(\ref{2.16}) and
(\ref{2.19}) introduced in Sect. 2.1, is based on the following two basic ingredients \cite{fadeev}
\begin{itemize}
\item zero curvature representation
\br
\pa \bar {\cal A} - \bar \pa  {\cal A} - [{\cal A},  \bar {\cal A}] =0, 
\quad {\cal A},  \bar {\cal A}\in \oplus_{i=0,\pm 1} \lie_i
\label{2.202}
\er
of their equation of motion.

\item fundamental Poisson bracket (FPR) relation
\br
\{{\cal A}_x(x_1,t) \stackrel{\otimes}{,} {\cal A}_x(x_2,t) \}_{PB} = [r_{cl}, \; {\cal A}_x(x_1,t) \otimes I + I \otimes {\cal A}_x(x_2,t) ] \d(x_1-x_2),
\label{2.203}
\er
where $r_{cl}$ denotes the classical $r$-matrix and ${\cal A}_x = {1\o 2} ({\cal A} - \bar {\cal A})$.

\end{itemize}

The Leznov-Saveliev matrix form \cite{lez-sav} of the equations of motion derived from the 
corresponding Lagrangians 
(\ref{1.9}),(\ref{2.13}),(\ref{2.16}) and (\ref{2.19}) is 
\br
\bar \pa ( g_0^{-1} \pa g_0 ) + [\eps_-, g_0^{-1} \eps_+ g_0 ] =0, \quad 
 \pa (\bar \pa g_0 g_0^{-1}) - [\eps_+, g_0 \eps_- g_0^{-1} ] =0,
\label{2.21}
\er
with $g_0 (\psi_a, \chi_a, \varphi_l, R_a)\in G_0$ given by eqns. (\ref{2.6}), (\ref{2.12}) and (\ref{2.172}). It
allows us to deduce the explicit form of the 2-d pure gauge potentials, namely, 
\br
{\cal A}=  - g_0 \eps_- g_0^{-1}, \quad \bar {\cal A} =\eps_+ + \bar \pa g_0 g_0^{-1}
\label{2.212}
\er
They indeed satisfy eqns. (\ref{2.202}) if $g_0$ is  a solution of eqns. (\ref{2.21}).  Their explicit form in terms of the
fields $\varphi_l, \psi_a, \chi_a, R_a $, etc  is given in refs. \cite{elek}, \cite{tau}, \cite{new},
\cite{wigner}.

The derivation of $r_{cl}$ and the proof of eqn. (\ref{2.203}) is based on the explicit realization of $({\cal A}_x)_{ij}$
in terms of fields and their canonical momenta.  We next calculate the equal time matrix PBs:
\br
\{{\cal A}_x(x_1,t) \stackrel{\otimes}{,} {\cal A}_x(x_2,t) \}_{ij;kl} =\{({\cal A}_x)_{ik}, ({\cal A}_x)_{jl} \}
\nonu
\er
making use of the basic canonical PBs, say $\{ \phi_j (x,t), \Pi_{\phi_k}(y,t)\} = \d_{jk}\d(x-y)$.  The result for the IM
(\ref{1.2}) has the form \cite{dubna} 
\br
r_{cl} = \b^2 \( C^{+} - C^{-} \)
\label{2.23}
\er
where
\br
C^{+} &=& \sum_{m=1}^{\infty}  \( \sum_{a,b=1}^{n} (K^{-1})_{ab} h_a^{(m)}\otimes h_b^{(-m)} + 
\sum_{\a >0} (E_{\a}^{(m)} \otimes  E_{-\a}^{(-m)} + E_{-\a}^{(m)} \otimes  E_{\a}^{(-m)})\) \nonu \\
&+&
{1\o 2} \sum_{\a >0} E_{\a}^{(0)} \otimes  E_{-\a}^{(0)},
\label{2.24}
\er
and $C^- = \s (C^+ ), \;\; \s (A\otimes B) = B \otimes A$; $K^{-1}_{ab}$ denotes the inverse of the Cartan matrix for
$\lie A_n$.  It turns out that $r_{cl}$ has an universal form (\ref{2.23}), (\ref{2.24}) for each given algebra, say,
$A_n^{(1)}$ and does not depend on the choice of the specific graded structure and  of $\eps_{\pm}$, i.e., 
 the abelian and nonabelian affine $A_n$-Toda model provide different representations of the same  
 FPR algebra (\ref{2.203}) \footnote{ It is not however true for the quantum $R$-matrices, which acquire specific form for
 each fixed graded structure and $\eps_{\pm}$}.

The complete proof of the integrability of the considered models require two more steps.  The first is the construction of
the infinite set of conserved charges.  Their existence and explicit form is well known consequence of the zero curvature
representation (\ref{2.202}) \cite{fadeev}, namely
\br
Tr ( T(\tau )^m) &=& P_m(\tau ), \quad \pa_{\tau} P_m = 0 \;\;\;\; m=0, \pm 1, \cdots \nonu \\
T(\tau )&=&    \lim \limits_{L \to \infty} P \exp \int_{-L}^{L} {\cal A}_x (\tau, x)dx
\label{2.25}
\er
i.e., $P_m(0)$ are the conserved charges we seek.  The last step is to prove that these conserved charges are in involution,
i.e.
\br
\{ P_{m_1},P_{m_2} \}_{PB} = 0
\nonu
\er
which is ensured by the specific form of the FPR (\ref{2.203}) and the explicit form of the charges $P_m$ (\ref{2.25}).

\subsection{Identification of Dyonic IMs as perturbed CFTs}

The Hamiltonian reduction \cite{or} and the integrable relevant perturbations of certain
 conformal minimal models \cite{z1},\cite{z2},\cite{z3},
are known to be two equivalent methods for constructing and solving a large class of IMs.
  As we have shown in Sect. 2.1,
the first method is based on an appropriate graded structure ($Q, G_0, \eps_{\pm}$) of the 
defining affine algebra $\hat
\lie $ (say, $A_n^{(1)}$) and each choice of $\lie, Q$ and $\eps_{\pm}$ determines one IMs.  The second method introduced by
Zamolodchikov \cite{z1}, \cite{z3}, 
consists in considering  certain quantum CFT (i.e. conformal minimal model of one 
of the extended Virasoro
algebras: conformal current algebra, $W_n$, $Z_n$-parafermionic algebra, \cite{pf}, \cite{fz1}, etc.) and adding to its Lagrangian a linear
combination $\sum_{a=1}^{l} g_aV_a\bar V_a$ of certain {\it relevant} vertex operators of dimension $\Delta_a + \bar
\Delta_a<2$ representing highest weight representations of the correseponding extended conformal algebra, i.e.,
\br
{\cal L}_{pert} = {\cal L}_{conf} + \sum_{a=1}^{l} g_a V_a(z) \bar V_a(\bar z)
\label{2.27}
\er
The relation between these two methods has been demonstrated on the examples of sine-Gordon (SG) model  as $\phi_{1,2} \bar
\phi_{1,2}$ perturbation of the Virasoro minimal models \cite{s}, abelian affine Toda as $W_{n+1}$ minimal model
perturbations \cite{vega}, the Lund-Regge model as $\phi_{0,0}^{(2)}$ perturbation of the $Z_N$ parafermions \cite{bakas},
\cite{f}, etc. (see ref. \cite{rev} for review). The problem  we address in this Section is to 
recognize the IMs derived in Sect.2.1 as perturbed CFTs. 
Therefore we have to answer the following two questions:
\begin{itemize}

\item  which are the extended conformal algebras behind the conformal limits $V=V_{conf} +
 V_{pert} \rightarrow V_{conf}$ of
the dyonic IMs (\ref{1.2}), (\ref{1.9}),(\ref{2.13}),(\ref{2.16}) and (\ref{2.19})?

\item how to identify the nonconformal part of their potentials $V_{pert}$, say for the IM (\ref{1.9})
\br
V_{pert}^{(g)} = e^{\b (\varphi_1 + \varphi_{n-1})} (1+ \b^2 \psi_g \chi_g e^{-\b \varphi_1})
\label{2.28}
\er
with certain linear combinations of vertex operators of the underlying conformal models (of the 
$V_{n+1}$-algebra \cite{annal1}, \cite{pl}, \cite{quantum}
for the gauged IM (\ref{1.9}))?

\end{itemize}

One should consider separately the $U(1)$-IMs (\ref{1.2}) (and (\ref{1.9})) with $n=1$ and $n \geq 2$, due to the fact that
they are based on different type of extended conformal algebras, namely:

\begin{itemize}
\item the $n=1$ ungauged IM (\ref{1.2}) is governed by the chiral $SL(2,R)$ conformal current algebra and its gauged version
$SL(2,R)/U(1)$ - by the parafermionic algebra \cite{pf}.

\item the $n \geq 2$ ungauged IMs (\ref{1.2}) appear to be certain perturbations of the 
minimal models of the
$W_{n+1}^{(n)}$-Bershadsky-Polyakov algebra \cite{ber}, \cite{pol}. 
 The conformal limit of the gauged IMs (\ref{1.9}) with 
$n \geq 2$  is characterized by the nonlocal $V^{(1)}_{n+1}$-algebra of mixed PF-$W_n$-type
 \cite{annal1}, \cite{pl}, \cite{quantum}.

\end{itemize}

Similar separation takes place in the case of $U(1)\otimes U(1)$ multicharged IMs (\ref{2.13}),
 (\ref{2.16}) and 
(\ref{2.19}).  The simplest case $n=2$ represents relevant perturbations of the $SL(2,R)\otimes SL(2,R)$ 
 WZW models, while the $n\geq 3$ case is related to specific quadratic (non-Lie)
$W_{n+1}^{(n,2)}$-algebra spanned by two sets $J^{\pm}_{\a_a}, a=1,n$ of spin $s= {n\o 2}$ currents,
 two spin $s=1$ currents
$J_{\l_a\cdot H}$ and $n-2$ currents $T_s$ of spin $s=2,3, \cdots n-1$.  Its algebraic structure (i.e.
 the OPE's of these currents) 
is quite similar to the one
of $W_{n+1}^{(n)}$ algebra, but including an extra set of currents $J^{\pm}_{\a_n}, \;\; J_{\l_n\cdot H}$.

We first consider few different integrable perturbations of the $SL(2,R)$-WZW model \cite{bakas}, \cite{braz},
\cite{bonora}. 
 For $n=1$ the Lagrangian
(\ref{1.2}) takes the form
\br
{\cal L}_u^{p=1} (n=1)= \pa R \bar \pa R + \pa \chi_u \bar \pa \psi_ue^{2\b R} - V_u, 
\quad V_u = m^2 \chi_u \psi_u e^{2\b R}, \quad R = {1\o 2}R_u
\label{2.29}
\er
Its conformal limit $V_u \rightarrow 0$, (i.e. $m^2 \rightarrow 0$ ) coincides with the $SL(2,R)$-WZW model.  We next
remember that the  vertex operators, representing the primary fields of the discrete series of highest weight
representations \cite{kz}, \cite{wak}, \cite{ger} are given by 
\br
V_{m, \bar m}^j(z, \bar z) = \psi^{j+m} \chi^{j+\bar m} e^{2\b jR} \equiv \Phi_j^m (z) \bar \Phi_j^{\bar m}(\bar z), \quad m,
\bar m =-j,-j+1,  \cdots , j
\label{2.30}
\er
of conformal dimensions $\Delta_{\Phi} = \bar \Delta_{\bar \Phi} ={{j(j+1)}\o {k-2}}$, 
i.e. $\Delta_V = 2\Delta_{\Phi}$. 
Therefore the perturbation of $SL(2,R)$  WZW model by $V_{0,0}^{(1)}$, can be represented
 by ${\cal L}_u^{p=1}(n=1)$ of eqn.
(\ref{2.29}).  Since $V_u = Tr \eps_+g_0\eps_-g_0^{-1}- Tr \eps_+\eps_- = V_{0,0}^{(1)}$ 
for $\eps_{\pm} = mh^{(\pm 1)}$ of
grade $\pm 1$ with respect to the homogeneous grading $Q=d$ and $g_0 \in SL(2,R)$ we conclude 
that the IM ($n=1$) defined
by the above grading structure $Q, \eps_{\pm}$ of $\hat {G}_0 = SL(2,R)$ is identical with the 
$V_{0,0}^{(1)}$ perturbation
of the $SL(2,R)$ WZW model.  Note that in the classical limit $k \rightarrow \infty $ and 
$\Delta_V \rightarrow 0$ and
therefore the constant $m$ has dimension $1$ in mass units.  If we take the most general grade $\pm 1$ elements
\br
\eps_{\pm} = m \( a_1^{(\pm)} h^{(\pm 1)} + a_2^{(\pm)} E_{\a}^{(\pm 1)} + a_3^{(\pm)} E_{-\a}^{(\pm 1)} \)
\label{2.31}
\er
instead of $\eps_{\pm} \sim h^{(\pm 1)}$, we get a six parameter family of integrable perturbations of the 
$SL(2,R)$ WZW model
\br
V_u &=& ({1\o 2} a_1^+ + a_3^+ \psi )({1\o 2} a_1^- + a_2^- \chi ) + a_2^- a_3^+ e^{-2\b R} \nonu \\
&+& 
(a_1^+ \psi + a_3^+\psi^2 - a_2^+ ) (a_1^- \chi + a_2^-\chi^2 - a_3^- )e^{2\b R}
\label{2.32}
\er
considered in ref. \cite{braz}.  For arbitrary $a^{\pm}_i$ both $SL(2,R)_{left}$ and $SL(2,R)_{right}$ are broken. 
 Moreover, when $a_i^+ =a_i^- =a_i$, we have
\br
[a_1h^{(0)} + a_2E_{\a}^{(0)} + a_3E_{-\a}^{(0)}, \eps_{\pm }]=0
\nonu
\er
and therefore such perturbations has chiral $U(1)_{left}\otimes U(1)_{right}$
symmetries.  Particular examples of chiral integrable perturbations by
\br
V_u^{\pm} = \Phi_j^{\pm j}(z), \quad V_u^{(0)} = \Phi_j^0(z)
\nonu
\er
have been introduced and studied in ref. \cite{bonora}.  It is important to mention that exhausting the possible gradings
($Q=d$) and $\eps_{\pm}$'s (see eqn. (\ref{2.31})) one can classify all the integrable perturbations of a given 
WZW model, i.e. listing the admissible graded structures ($\hat G, Q, \eps_{\pm}$) one separates few integrable linear
combinations of the vertices $V^j_{m, \bar m}$ among the large set of combinations with $j\leq {k \o 2}$.

In the cases when the perturbation preserves the chiral $U(1)$ symmetry (i.e. $a_i^{+} =a_i^-$)
 one can further gauge fix
this symmetry as it was explained in Sect. 2.1 (see eqn. (\ref{2.8})).  The corresponding $n=1$
 gauged IMs (\ref{1.9}) (or
more general for $\eps_{\pm}$ given by eqn. (\ref{2.31})) give rise to different integrable 
perturbations of the gauged
$SL(2,R)/U(1)$ - WZW (i.e. noncompact parafermions \cite{dixon}) and $SU(2)/U(1)$ - (i.e. compact parafermions) 
studied in ref. \cite{f}, \cite{bakas}.

The conformal limits of IMs (\ref{1.2}) with $n\geq 2$ represent certain conformal $A_n$-non-abelian 
Toda models introduced
in ref. \cite{annal1}, \cite{pl} (see Sect. 2 and 8 of ref. \cite{annal1}).  
They can be defined as conformal gauged 
$G_0 ={ H}_{-}\backslash A_n / { H}_{+}$-WZW model based on the finite dimensional Lie algebra $A_n$, with  
$H_{\pm} \in A_n$ being  the  of positive and negative graded nilpotent subalgebras
 according to the  following grading operator
 \br
Q^{conf} = \sum_{i=2}^{n} \l_i\cdot H^{(0)} , \quad \eps_{\pm}^{conf} = \sum_{i=2}^{n} E_{\pm \a_i}^{(0)},
\label{2.33}
\er
Similar to the conformal abelian Toda theory, the nontrivial conformal part \\
$V_{conf} = Tr (\eps_+^{conf} g_0 \eps_-^{conf}
g_o^{-1})$ of potential $V_u (n\geq 2)$ is originated from the specific set of constraints on the $A_n$-WZW currents:
\br
J_{-\a_i} = \bar J_{\a_i} = 1, \quad i=2, \cdots n, \quad J_{-[\a ]} = \bar J_{[\a ]} = 0, [\a ] = {\rm non \;\; simple \;\;
root}
\label{2.33a}
\er
encoded in $\eps_{\pm}^{conf}$.  
The vertices $V_j = e^{-\b \sum \eta_{ij} \varphi_i }$ of dimension (1,1) 
represent the screening operators of the
$W_{n+1}^{(n)}$-algebra.  As it is well known these constraints reduce the
 original $A_n$-chiral conformal current algebra
to specific higher spin quadratic algebra of $W_{n+1}^{(n)}$-type studied
 by Polyakov \cite{pol} and Bershadsky \cite{ber}.  For
example the $n=2$ Bershadsky-Polyakov (BP) algebra $W_3^{(2)}$ is generated by four chiral 
currents $T_{W}, G^{\pm}$ and $J$ of spins
$s=2,{3\o 2},{3\o 2}, 1$.  It has the following OPE form \cite{ber}
\br
J(z_1) J(z_2) &=& {{2k+3}\o {3z_{12}^2}} + O(z_{12}), \quad G^{\pm}(z_1) G^{\pm}(z_2) = O(z_{12}) \nonu \\
J(z_1) G^{\pm} (z_2) &=& \pm {1\o {z_{12}}}G^{\pm} (z_2)+ O(z_{12}) \nonu \\
T_W(z_1) T_W(z_2) &=& {{c_W}\o {2z_{12}^4}} + {2\o {z_{12}^2}}T_W(z_2) + {1\o {z_{12}}}\pa T_W(z_2) + O(z_{12})\nonu \\
T_W(z_1) G^{\pm}(z_2) &=&{3\o {2z_{12}^2}}G^{\pm}(z_2) + {1\o {z_{12}}} \pa G^{\pm}(z_2) + O(z_{12})\nonu \\
T_W(z_1) J(z_2) &=& {1\o {z_{12}^2}} J(z_2) + {1\o {z_{12}}} \pa J(z_2)+ O(z_{12})\nonu \\
G^{+}(z_1) G^{-} (z_2) &=&(k+1) {{(2k+3)}\o {z_{12}^3}} + 3 {{(k+1)}\o {z_{12}^2}}J(z_2) \nonu \\
&+& {1\o {z_{12}}}\( 3:J^2: -
(k+3) T(z_2) + {3\o 2} (k+1) \pa J \) + O(z_{12})
\label{2.35}
\er
where $c_W = {{8k}\o {k+3}} -6k-1$ is its central charge.
In fact the original constraints of the $n=2$ BP model 
\br
J_{-\a_2} = \bar J_{\a_2}=0, \quad J_{-\a_1-\a_2} = \bar J_{\a_1+\a_2}=1
\nonu
\er
are the image of the (\ref{2.33a}) under the action of particular $A_2$-Weyl reflection $w_{\a_1}$, i.e.,
\br
w_{\a_1}(\a ) =\a - (\a \cdot \a_1 ) \a_1, \quad w_{\a_1}^2 =1.
\nonu 
\er
as it has been shown in Sect 8 of ref. \cite{annal1}.  According to the analysis of the h.w. representations of the $W_3^{(2)}$
algebra (\ref{2.35}) one can construct a specific class of degenerate vertex operators, that realize in the $\psi_u, \chi_u,
R_u, \varphi$ variables of our model (\ref{1.2}) acquire the form:
\br
V_{m, \bar m}^{j, \a_{r,s}, \b_{r,s}} (z, \bar z) = \psi_u^{j+m} \chi_u^{j+\bar m} e^{\b (\b_{r,s} R_u + \a_{r,s}\varphi )}
\label{2.36}
\er
where $\a_{r,s}$ and $ \b_{r,s}$ are {certain charges defined in Sect. 9 of ref. \cite{annal1} (see also
\cite{ber} })
For $j=1,\;  m=\; \bar m =0$, $\a_{r,s} = \b_{r,s} =1$ and for  $j= m = \bar m = \b_{r,s}=0, \; \a_{r,s} =2$ one recognize
the two vertex operators that form the nonconformal part (i.e., the integrable perturbation) $V_{pert}$ of $V_u$, i.e.,
\br  
 V_{pert}= e^{2\b  \varphi } + \b^2 \psi_u \chi_u e^{\b(R+\varphi )} = V_{0,0}^{0,2,0} + \b^2 V_{0,0}^{1,1,1}
\nonu
\er
Similar identifications takes place for the gauged version (\ref{1.9}) of the $n=2$ IM (\ref{1.2}). 
 In this case the role
of the $W_3^{(2)}$ algebra and its vertex operators (\ref{2.36}) is played by the $V_3^{(1,1)}$ algebra
 of mixed
PF-$W_2$-type \cite{annal1}, \cite{quantum} (spanned by the strees tensor $T_v$ and two PF currents $V^{\pm}$ of spin 
$s^{\pm} = {3\o 2}(1- {{1}\o
{2k+3}})$) and the corresponding h.w. vertex operators.  The conformal extended algebras $V^{(1,1)}_{n+1}$ of the classical
symmetries of the conformal limits $V_{pert} \rightarrow 0$ for generic $n\geq 2$ gauged IMs (\ref{1.9}) have been
constructed in ref. \cite{annal1}.  Their quantum versions, the h.w. representations and the vertex operators are known,
however only in the particular case of $n=2$, (i.e. $V_3^{(1,1)}$).  Therefore, the problem of identification of $n\geq 3$
gauged (and ungauged) dyonic IMs as perturbed CFT m.m's of the $V_{n+1}^{(1,1)}$ (and $W_{n+1}^{(n)}$) algebras remains
open.

\sect{Dyonic IMs with $U(1)$ local symmetry}

\subsection{Gauged vs. ungauged IMs: conserved charges relations}

The invariance of the ungauged IM Lagrangian (\ref{1.2}) under local $U(1)$ transformations (\ref{1.3}) (see eqn.
(\ref{2.7}) for their matrix form) gives rise to the following chiral $U(1)$ conserved currents:
\br
J(z)&=& Tr (g_0^{-1} \pa g_0 \l_1 \cdot H^{(0)}) = {{2n}\o {n+1}}\b_0 \pa R_u - 2i \b_0^2 \psi_u \pa \chi_u
e^{i\b_0(R_u-\varphi_1)}, \nonu \\
\bar J(z)&=& Tr (\bar \pa g_0 g_0^{-1}\l_1 \cdot H^{(0)}) = {{2n}\o {n+1}}\b_0 \bar \pa R_u - 2i \b_0^2 \chi_u \bar \pa \psi_u
e^{i\b_0(R_u-\varphi_1)}
\label{3.1}
\er
i.e., $\bar \pa J = \pa \bar J =0$ as one can verify by taking traces of eqns. (\ref{2.21}) 
with $\l_1 \cdot H^{(0)}$ (remembering
that $[\l_1 \cdot H^{(0)}, \eps_{\pm}]=0$).  Hence the solutions of eqns. (\ref{2.21}) are characterised by 
two infinite sets of
conserved charges ($m\in Z$),
\br
Q_{m+1} = \oint J(z)z^{-m-1} dz, \quad \bar Q_{m+1} = \oint \bar J(\bar z)\bar z^{-m-1} d\bar z
\label{3.2}
\er
Since the conservation of the chiral $U(1)$ currents $J$ and $\bar J$ is a consequence of the conservation of both vector
$J_{\mu}^{vec} = J_{\mu}$ and axial $J_{\mu}^{axial} = \eps_{\mu \nu}J^{\nu}$ currents (i.e., $\pa^{\mu}J_{\mu}^{vec,
axial}=0$ and therefore $J_{\mu} = \pa_{\mu} \Phi, \quad \Phi = \Phi(z) + \bar \Phi (\bar z) $)  we realize that 
\br
J=J_0 +J_1 = \pa \Phi (z), \quad \bar J=J_0 - J_1 =  \bar \pa \bar \Phi (\bar z)
\label{3.3}
\er
It implies the following relations between axial and vector $U(1)$-charges $Q_{vec}, Q_{axial}$ with the $J, \bar J$-zero
modes $Q_{0}, \bar Q_{0}$:
\br
Q_{vec} &=& \int_{-\infty}^{\infty} J_0 dx = {1\o 2} \int_{-\infty}^{\infty}(J + \bar J)dx = {1\o 2} (Q_0 + \bar Q_0), \nonu
\\
Q_{ax} &=& \int_{-\infty}^{\infty} J_1 dx = {1\o 2} \int_{-\infty}^{\infty}(J - \bar J)dx = {1\o 2} (Q_0 - \bar Q_0)
\label{3.4}
\er
Taking into account the ``chirality'' of the free fields $\Phi(z), \bar \Phi(\bar z)$, i.e., $\bar \pa \Phi(z) = 
\pa \bar \Phi(\bar z)= 0$, we have $ \pa \Phi = 2 \pa_x \Phi$ and $ \bar \pa \bar \Phi =-2 \pa_x \bar \Phi $. Therefore
all charges $Q_{vec}, Q_{axial}, Q_0, \bar Q_0$ are defined in terms of the asymptotics of $\Phi$ and $\bar \Phi$ at $x
\rightarrow \pm \infty $ (and fixed $t$, say $t=0$):
\br
Q_0 = 2\int_{-\infty}^{\infty}\pa_x \Phi dx = 2 \( \Phi(\infty ) - \Phi(-\infty )  \), \nonu \\
\bar Q_0 = -2\int_{-\infty}^{\infty}\pa_x \bar \Phi dx = -2 \( \bar \Phi(\infty ) - \bar \Phi(-\infty )  \)
\label{3.5}
\er
We next observe that the second terms of $J$ and $\bar J$ given by eqn. (\ref{3.1}) are in fact components of the global
$U(1)$-current $I_{\mu}^u = (I,\bar I)$, i.e.,
\br
I^u &=& 2i\b_0^2 \psi_u \pa \chi_u e^{i\b_0(R_u -\varphi_1)}, \nonu \\
\bar I^u &=& -2i\b_0^2 \chi_u \bar \pa \psi_u e^{i\b_0(R_u -\varphi_1)}, \quad \bar \pa I^u + \pa \bar I^u =0,
\label{3.6}
\er
generated by the following global $U(1)$-transformations ($\eps$= const):
\br
\psi^{\pr}_u = \psi_u e^{i\b_0^2 \eps}, \quad  \chi^{\pr}_u = \chi_u e^{-i\b_0^2 \eps}, \quad R^{\pr}_u = R_u, \quad
\varphi_l^{\pr} = \varphi_l
\label{3.7}
\er
As one might expect, its charge  $Q^{el}_u = {1\o 2} \int_{-\infty}^{\infty} (I+ \bar I)dx$ is not independent of $Q_0$ and
$\bar Q_0$:
\br
Q_u^{el} = {{2n}\o {n+1}} Q_R^u - {1\o 2} (Q_0 - \bar Q_0), \quad \quad Q_R^u = \b_0 \int_{-\infty}^{\infty} \pa_x R_u dx
\label{3.8}
\er
due to the relation 
\br 
{1\o 2} (J- \bar J) = {{2n \b_0}\o {n+1}} \pa_x R_u - {1\o 2} (I+\bar I)
\label{3.82}
\er
between the currents $J, \bar J, I, \bar I$ and the topological current $J_{R_u}^{\mu} = \b_0 \eps^{\mu \nu} \pa_{\nu}
R_u$, encoded in eqn. (\ref{3.1}).  It reflects the fact that the transformation (\ref{3.7}) is a particular case $w=\bar w
= -\eps $ of eqn. (\ref{1.3}). 

As we have shown in Sect. 2.1, to each ungauged $G_0$-IM with local  $G_0^0 \subset G_0$ symmetry, one can make in
correspondence a new  $G_0/G_0^0$-gauged IM, by gauge fixing $G_0^0$.  The procedure of elimination of the extra 
$G_0^0$-field degrees of freedom ($R_u$  for $G_0^0 = U(1)$) consists in imposing constraints on the $G_0^0$-chiral currents
($J(z)$ and $\bar J(z)$):
\br
J_g &=&Tr ( (g_0^f)^{-1}\pa g_0^f \l_1\cdot H) \nonu \\
&=& {{2n}\o {n+1}} \b_0 \pa R_g (1+ \b_0^2 {{n+1}\o {2n}}\psi_g\chi_g e^{-i\b_0
 \varphi_1}) -2i\b_0^2 \psi_g\pa \chi_g e^{-i\b_0 \varphi_1} = 0 \nonu \\
\bar J_g &=&Tr ( \bar \pa g_0^f (g_0^f)^{-1}\l_1\cdot H) \nonu \\
&=& {{2n}\o {n+1}} \b_0 \bar \pa R_g 
(1+ \b_0^2 {{n+1}\o {2n}}\psi_g\chi_g e^{-i\b_0
\varphi_1}) -2i\b_0^2 \chi_g\bar \pa \psi_g e^{-i\b_0\varphi_1} = 0 
\label{3.9}
\er
The relation between the field variables $g^u_0 (\psi_u, \chi_u, R_u, \varphi_l) \in G_0$ of 
the ungauged IM (\ref{1.2}) 
\br
g_0^u = e^{\chi_u E_{-\a_1}^{(0)}} e^{\b (\l_1\cdot H^{(0)}R_u + \sum_{i=1}^{n-1}\varphi_i h_{i+1}^{(0)})} e^{\psi_u E_{\a_1}^{(0)}}
\nonu 
\er
and of the gauged IM (\ref{1.9}) $g^f_0 (\psi_g, \chi_g,  \varphi_l) \in G_0/G_0^0$ 
\br
g_0^f = e^{\chi_g E_{-\a_1}^{(0)}} e^{\b \sum_{i=1}^{n-1}\varphi_i h_{i+1}^{(0)}} e^{\psi_g E_{\a_1}^{(0)}}
\nonu 
\er
is given by 
\br
g_0^u = e^{\b \bar w(\bar z)\l_1\cdot H^{(0)}} \a_0 g_0^f \a_0 e^{\b  w( z)\l_1\cdot H^{(0)}}, 
\quad \a_0 = e^{{1\o 2}\b \l_1\cdot H^{(0)}R_g}
\label{3.10}
\er
In   components we find the following relations:
\br
R_u &=& R_g +\b_0 (w+\bar w), \quad \varphi_l^u = \varphi_l^g = \varphi_l, \nonu \\
\psi_u &=& \psi_g e^{-{1 \o 2}i\b_0 R_g - i \b_0^2 w}, \quad \chi_u = \chi_g e^{-{1 \o 2}i\b_0 R_g - i \b_0^2 \bar w}, 
\quad \theta_g = {{1}\o {2i \b_0}} ln (\chi_g/ \psi_g ) 
\nonu \\
\theta_u &=& \theta_g - {1\o 2}\b_0 (w-\bar w), \quad \theta_u = {{1}\o {2i \b_0}} ln (\chi_u/ \psi_u )
\label{3.11}
\er
They reflect (a) the axial gauge fixing of chiral $U(1)_L\otimes U(1)_R$ symmetry and (b) the specific choice of the
representative $g_0^f$ of the coset $G_0/G_0^0$ (i.e. of the nonlocal field $R_g$) such that the ${\cal L}_{p=1}^{g}$
(\ref{1.9}) is local in $\psi_g, \chi_g$, i.e., independent of $R_g$.  It is instructive to verify that by gauge
transforming $R_g, \psi_g$ and $\chi_g$ according to (\ref{3.11}) with 
\br
w={{1}\o {\b_0^2}} \({{n+1}\o {2n}}\) \Phi, \quad \quad \bar w={{1}\o {\b_0^2}} \({{n+1}\o {2n}}\) \bar \Phi
\nonu
\er
one recover eqn. (\ref{3.1}) starting from the constraints (\ref{3.9}) and vice versa.

The form of the global $U(1)$ current $I_{\mu}^g = (I^g, \bar I^g)$ of the gauged IM (\ref{1.9}) 
\br
I^g = 2i\b_0^2 {{ \psi_g \pa \chi_g e^{-i\b_0 \varphi_1}} \o
{1+ \b_0^2 {{n+1}\o {2n}} \psi_g\chi_g e^{-i\b_0  \varphi_1}}}, \quad  \quad 
\bar I^g = -2i\b_0^2 {{ \chi_g \bar \pa \psi_g e^{-i\b_0  \varphi_1}} \o
{1+ \b_0^2 {{n+1}\o {2n}} \psi_g\chi_g e^{-i\b_0 \varphi_1}}}
\label{3.12}
\er
i.e., $\bar \pa I^g + \pa \bar I^g =0$, suggests that the constraints (\ref{3.9}) (i.e. the gauge 
fixing conditions) can be
considered as a requirement of the proportionality of the electric current $I^g_{\mu}$ (\ref{3.12}) and the
$R_g$-topological current $J^{R_g}_{\mu} = \b_0 \eps_{\mu \nu } \pa^{\nu} R_g$, i.e.,
\br
I^g_{\mu} = {{2n}\o {n+1}} J^{R_g}_{\mu} 
\label{3.13}
\er
Note that in the  ungauged IM (\ref{1.2}) this relation is replaced by the more general one (\ref{3.82})
\br
I^u_{\mu} = \eps_{\mu \nu }\( {{2n}\o {n+1}} \pa^{\nu} R_u - J^{\nu} \) 
\label{3.14}
\er
and therefore $I_{\mu}^u$ and $J^{R_u}_{\mu}$ are independent due to the contribution of $J^{\nu}$. 
 The relation
(\ref{3.11}) establised between the fields of the gauged and ungauged IMs allows us to relate the 
corresponding $U(1)$ and
topological currents and charges of both models:
\br
Q_R^u &=& Q_R^g + {{n+1}\o {4n}}(Q_0 -\bar Q_0) = Q_R^g + {{n+1}\o {2n}} Q_{ax}
\nonu \\
Q_{\theta}^u &=& Q_{\theta}^g + {{n+1}\o {8n}}(Q_0 +\bar Q_0) = Q_{\theta}^g + {{n+1}\o {4n}} Q_{vec}
\label{3.15}
\er
where we have introduced the topological charge $Q_{\theta}$ related to the field $\theta_g = {{1}\o {2i \b_0}}ln (\chi_g
/\psi_g )$.  Replacing $Q_R^u $ in eqn. (\ref{3.8}) we find that 
\br
Q_u^{el} = {{2n}\o {n+1}}Q_R^g = Q_g^{el}
\label{3.16}
\er
i.e., the global $U(1)$ charges (of the currents $I_u^{\mu}$ and $I_g^{\mu}$) do coincide.  This leads to the conclusion
that one can determine the charge spectrum of the ungauged IM in terms of the charges $Q_R^g, Q_{\theta}^g, Q^{el}_g$ (and 
$Q_{top}^{\varphi_l} = \b_0 \int_{-\infty}^{\infty}\pa_x \varphi_l dx$) of the gauged IM and the $J, \bar J$ charges
($Q_{vec}, Q_{ax}$ or $Q_0, \bar Q_0$), i.e., the asymptotics of $\Phi, \bar \Phi$.

\subsection{Nonconformal GKO coset construction}
 The relation between the fields ($g_0^u, g_0^f$ and $\Phi, \bar \Phi$), currents and charges (\ref{3.14}), 
(\ref{3.15}) and (\ref{3.16}) of the ungauged $G_0$-IM, the gauged $G_0/G_0^0$-IM and the chiral $U(1)$-CFT addresess the
question whether the stress-tensors (and the energies) of these theories are related in similar manner. Having in mind
that for the conformal limits\footnote{i.e., negleting the perturbations $V_u^{pert} = {{m^2}\o {\b^2}}\( e^{\b (\varphi_1
+ \varphi_{n-1})} (1+\b^2 \psi_u \chi_u e^{\b (R_u - \varphi_1)})-n \)$ and $ V_g^{pert}$}
of the considered IMs (\ref{1.2}) and (\ref{1.9}) according to Goddard-Kent-Olive (GKO) coset 
construction \cite{god}, we have
\br
T_{G_0}^{CFT} = T_{G_0/G_0^0}^{CFT} + T_{G_0^0}^{CFT}
\label{3.17}
\er
one might expect that an appropriate nonconformal extension of the GKO formula (\ref{3.17}) to take place.

We start the derivation of the integrable models analog of eqn. (\ref{3.17}) by calculating  the ungauged IM stress-tensor
\br
T^u = {1\o 2}\eta_{ij} \pa \varphi_i  \pa \varphi_j + {{n}\o {2(n+1)}} (\pa R_u)^2   + 
\pa \chi_u \pa \psi_u e^{i\b_0 (R_u- \varphi_1)}+ V_u \label{3.18} \\
\bar T^u = T^u (\pa \rightarrow \bar \pa ), \quad \quad T^u_{00} = {1\o 2} (T^u + \bar T^u), \quad 
T^u_{01} = {1\o 2} (T^u - \bar T^u)
\nonu
\er
We next substitute $R_u, \psi_u$ and $\chi_u$ in $T^u$, taking into account eqns. (\ref{3.11}) and the constraints
(\ref{3.9}) as well.  By straightforward simplifications we realize that $T^u$ can be written in terms of the gauged IM
fields $\psi_g, \chi_g $ and $\varphi_l$ together with the  $U(1)$-CFT currents  $J= \b_0^2 {{2n}\o {n+1}} \pa w $ and 
$\bar J= \b_0^2 {{2n}\o {n+1}} \bar \pa \bar w$ only, i.e., 
\br
T^u = {1\o 2} \eta_{ij} \pa \varphi_i \pa \varphi_j + {{\pa \chi_g \pa \psi_g }\o {\Delta}}e^{-\b \varphi_1}
+ V_g + {{\b_0^2 n}\o 2(n+1)} (\pa w)^2
\label{3.19}
\er
where $\Delta = 1+ \b^2 {{n+1}\o {2n}}\psi_g \chi_g e^{-\b \varphi_1}$.
Finally, we remind that the stress-tensor of the gauged IM, derived from its Lagrangian (\ref{1.9}) has the form
\br
T^g = {1\o 2} \eta_{ij} \pa \varphi_i \pa \varphi_j +{{\pa \chi_g \pa \psi_g }\o {\Delta}}e^{-\b \varphi_1}
+ V_g 
\nonu
\er
Therefore the {\it nonconformal } version of the GKO formula (\ref{3.17}) is given by
\br
T_{G_0}^{u} = T_{G_0/G_0^0}^{g} + T_{G_0^0}^{CFT}, \quad \quad 
\bar T_{G_0}^{u} = \bar T_{G_0/G_0^0}^{g} + \bar T_{G_0^0}^{CFT}
\label{3.20}
\er
or equivalently 
\br
T_{00}^{u} = T_{00}^{g} + T_{00}^{CFT}, \quad \quad 
\bar T_{00}^{u} = \bar T_{00}^{g} + \bar T_{00}^{CFT}
\label{3.21}
\er
where $T^{CFT}, \bar T^{CFT}$ are the $U(1)$-CFT stress-tensors:
\br
T^{CFT}= {{\b_0^2 n}\o 2(n+1)} (\pa w)^2, \quad \quad 
\bar T^{CFT}= {{\b_0^2 n}\o 2(n+1)} (\bar \pa \bar w)^2
\label{3.22}
\er
Note that the $T^{CFT}, \bar T^{CFT}$ are chiral, i.e., $\bar \pa T^{CFT}= \pa  \bar T^{CFT}=0$ but $T^u, T^g, 
\bar T^u, \bar T^g $ are not, since the corresponding gauged and ungauged IMs are not conformal invariant.

The formal spliting of the $G_0$-ungauged IM fields, currents and stress-tensor in $U(1)$-CFT and gauged $G_0/G_0^0$-IM
parts cannot  be considered as an indication of a direct sum,  since certain properties of the $U(1)$-CFT (b.c.'s of $w,
\bar w$) depend on the gauged IM b.c.'s  in the way indicated by the interaction
 terms of the ungauged IM, as we shall show in  Subsect. 3.3.

\subsection{Vacua, boundary conditions and discrete symmetries}

The potential $V_u$ of the ungauged IM (\ref{1.9}) for imaginary coupling $\b = i\b_0$ manifest $n$-distinct zeroes:
\br
\varphi_l^{(N)} = {{2\pi }\o {\b_0}}{{lN}\o {n}}, \quad \psi_u \chi_u = 0, \quad  R_u = \b_0 a_R, \quad \theta_u = \b_0
a_{\theta}
\label{3.24}
\er
where $N=0, \pm 1, \cdots \pm (n-1) \;\; {\rm mod} \;\; n, \quad l=1, 2, \cdots n-1$ and  $a_R, a_{\theta}$ are 
real parameters.   They represent the constant vacua solutions (i.e. $E_{vac}=0$, all charges $Q_{all}^{vac}=0$) of the
eqn. of motion (\ref{2.21}).   The vacua values and the boundary conditions of the {\it massless} fields $R_u$ and $\theta_u$
remain undefined by the $V_u=0$ condition \footnote{due to the fact that $\pa \tilde V_u / \pa R_u = 
\pa \tilde V_u / \pa \theta_u = 0 $, (i.e. $R_u$ and $\theta_u$ are flat directions) as one can see by suitable change of
variables $\psi_u = \tilde \psi_ue^{-{1\o 2}\b R_u}, \chi_u = \tilde \chi_ue^{-{1\o 2}\b R_u}$ and 
$V_u \rightarrow \tilde V_u$}.  As usually the global symmetries of the model ($Z_2\otimes Z_n$ and $U(1)_{vector}\otimes
U(1)_{axial}$ in our case (\ref{1.2})) determine the complete vacua structure and allowed b.c.'s of all the fields.  The
$Z_n$-group that leave ${\cal L}_{p=1}^{u}$ (\ref{1.2}) invariant acts as follows:
\br
\varphi_l^{\pr} =\varphi_l + {{2\pi }\o {\b_0}}{{lN}\o {n}}, \quad  \quad R_u^{\pr} = R_u + {{2\pi }\o {\b_0}}  (s_R + {{N-q-\bar
q}\o {n}})\nonu \\
\psi_u^{\pr} = \psi_u e^{\pi i ({{2q}\o {n}}+s_1)}, \quad 
\chi_u^{\pr} = \chi_u e^{\pi i ({{2\bar q}\o {n}}+s_2)}, \quad 
\theta_u^{\pr} = \theta_u + {{\pi}\o {\b_0}}({{\bar q -q}\o {n}} + s_{\theta}),
\label{3.25}
\er
where $q, \bar q = 0, \pm 1, \cdots \pm (n-1)$ and  $s_1,s_2, 2s_R=s_1+s_2,  s_R, s_{\theta}=s_2-s_1$ are integers.  
The $Z_n$-charges
of fields $\psi_u, \chi_u$ and $e^{\b \varphi_l}$ are given by $q , \bar q$ and $N$ mod $n$ respectively .  
One can further  combine the above $Z_n$  with the CP-transformation ($Px =-x, \;\; P\pa = \bar \pa, \;\; P^2=1$):
\br
\varphi_l^{\pr \pr} = \varphi_l^{\pr}, \quad R_u^{\pr \pr} = R_u, \quad \psi_u^{\pr \pr} = \chi_u, \quad 
\chi_u^{\pr \pr} = \psi_u
\label{3.26}
\er
into the larger diedral group $D_n$, requiring that $\psi_u$ and $\chi_u$ are conjugated, i.e.,
\br
\bar q = n-q
\label{3.27}
\er
and $w(t-x) \leftrightarrow \bar w(t+x)$.  Indeed by considering only chiral (say, left) $U(1)$ transformations ($w \neq 0,
\bar w =0$) one breaks the CP-invariance ($\bar q =0$, but $q \neq n$) and eqn. (\ref{3.26}) does not take place in this
case.  Completing the discussion about the relation between global symmetries and the vacua solutions we have also to
mention the following global $U(1)_{vector}\otimes U(1)_{axial}$ symmetries of ${\cal L}_{p=1}^{u}$ (\ref{1.2}):
\begin{itemize}
\item vector $U(1)$
\br
\psi^{\pr}_u = e^{-i\b_0^2 a_{\theta}}\psi_u, \quad \quad \chi^{\pr}_u = e^{i\b_0^2 a_{\theta}}\chi_u, \nonu \\
R_u^{\pr} = R_u, \quad \varphi_l^{\pr} = \varphi_l, \quad \theta^{\pr}_u = \theta_u + \b_0 a_{\theta}
\label{3.28}
\er
\item axial $U(1)$
\br
\psi^{\pr \pr}_u = e^{-i{1\o 2}\b_0^2 a_{R}}\psi_u, \quad \quad \chi^{\pr \pr }_u = e^{-i{1\o 2}\b_0^2 a_{R}}\chi_u, \nonu \\
R_u^{\pr \pr } = R_u + \b_0 a_R, \quad \varphi_l^{\pr \pr } = \varphi_l, \quad \theta^{\pr \pr }_u = \theta_u 
\label{3.29}
\er
\end{itemize}
Therefore the allowed b.c.'s for the fields at $x\rightarrow \pm \infty$ are given by
\br
\varphi_l^{(N)}(\pm \infty ) &=& {{2\pi}\o {\b_0}}{{lN_{\pm}}\o {n}}, \quad \quad 
R_u(\pm \infty ) = \b_0 a_R^{\pm} + {{2\pi}\o {\b_0}} (s_R^{\pm}+ {{N_{\pm} -q_{\pm} - \bar q_{\pm}}\o {n}}), \nonu \\
\psi_u \chi_u (\pm \infty )&=&0,\quad \quad 
\theta_u(\pm \infty ) = \b_0 a_{\theta}^{\pm} + {{\pi}\o {\b_0}} (s_{\theta}^{\pm}+ {{\bar q_{\pm} -  q_{\pm}}\o {n}})
\label{3.30}
\er

Together with the vacua sector (defined by $D_n\otimes U(1)_{vector}\otimes U(1)_{axial}$), the ungauged IM (\ref{1.2}) (in
contrast to the gauged (\ref{1.9})) admits a new {\it conformal sector}.  Due to chiral $U(1)$ symmetry (\ref{1.3}) its
equations of motion (\ref{2.21}) have conformal (1-D string-like) solutions
\br
\varphi_l^{(N)} = {{2\pi}\o {\b_0}}{{lN}\o {n}}, \quad \psi_u \chi_u =0, \quad
\theta_u^{CFT} = {{\b_0}\o 2}(\bar w-w), \quad R_u^{CFT }= \b_0 (w + \bar w)
\label{3.31}
\er
with nonvanishing energy $E^{CFT} = \pm P^{CFT}\neq 0$ and charges ($Q_0,  \bar Q_0 \neq 0 $)
 (\ref{3.5}).  The complete description of the $U(1)$ CFT representing the conformal sector of the IM (\ref{1.2}), requires
 the knowledge of the b.c.'s for the free fields $\Phi = {{2n\b_0^2}\o {n+1}} w$ and 
$\bar \Phi = {{2n\b_0^2}\o {n+1}} \bar w$.  The question to be answered is whether one can uniquely determine the $\Phi,
\bar \Phi$ b.c.'s  from the relation (\ref{3.11}) in terms of the b.c.'s of the fields of the ungauged IM ($ \psi_u, \chi_u,
R_u, \varphi_l$ given by eqn. (\ref{3.30})) and  of the gauged ones, $\psi_g, \chi_g, \varphi_l$ (see ref.
 \cite{elek}).  We
first consider the $Z_n$-transformations of $w$ and $\bar w$ (and $\Phi, \bar \Phi$). 
 According to eqn. (\ref{3.11}) we
have,
\br
e^{i\b_0^2 w} = {{\psi_g}\o {\psi_u}}e^{-i{1\o 2}\b_0 R_g}, \quad 
e^{i\b_0^2 \bar w} = {{\chi_g}\o {\chi_u}}e^{-i{1\o 2}\b_0 R_g}
\nonu 
\er
and therefore the $w, \bar w$ properties are a consequence of the $\psi_u, \chi_u$-transformation (\ref{3.25}) and of the
$\psi_g, \chi_g, R_g$ ones \cite{elek}:
\br
\psi_g^{\pr} = \psi_g e^{i\pi ({N\o n} + \tilde s_1)}, \quad 
\chi_g^{\pr} = \chi_g e^{i\pi ({N\o n} + \tilde s_2)}, \quad R_g^{\pr} = R_g, \quad \tilde s_1 + \tilde s_2 = 2 \tilde s
\label{3.32}
\er
where $\tilde s_1,\tilde s_2 $ and $\tilde s$ are integers ( $R_g = {{n}\o {n+1}}R$ in the notation of ref. \cite{elek}).
  The result  
\br
w^{\pr} = w + {{\pi}\o \b_0^2}(s + {{N-2q}\o n}), \quad 
\bar w^{\pr} = \bar w + {{\pi}\o \b_0^2}(\bar s + {{N-2\bar q}\o n}), \;\; s=\tilde s_1- s_1, \;\; \bar s = \tilde s_2
-s_2
\label{3.33}
\er
is consistent with the $R_u$ and $\theta_u$ transformations (\ref{3.25}) under the identification 
\br
s =s_R + s_{\theta}, \quad \quad \bar s = s_R - s_{\theta}
\label{3.34}
\er
Note that the particular case of left movers, i.e., $\bar w =0 $ and $w \neq 0$ takes place when the $Z_n$-charge $\bar q$ of
$\chi_u$ is half of the $\varphi_1$-charge: $\bar q = {{N}\o 2}$ and $ \tilde s_2 = s_2$.

The transformation properties (\ref{3.33}) of $w, \bar w$ allow us to single out an important class of b.c.'s (at $x
\rightarrow \pm \infty $) for the $U(1)$-CFT, namely, 
\br
\Phi(\pm \infty ) = {{2\pi }\o {n+1}} (ns_{\pm} + N_{\pm} - 2 q_{\pm }), \quad 
\bar \Phi(\pm \infty ) = {{2\pi }\o {n+1}} (n\bar s_{\pm} + N_{\pm} - 2 \bar q_{\pm })
\label{3.35}
\er
 They give rise to {\it topological} CFT solitons of vortex type constructed in the next Sect. 3.4.  Observe that for these CFT
 solutions with b.c.'s (\ref{3.35}) (i.e. interpolating between two coformal vacua  $s_+, q_+, N_+ \rightarrow 
s_-, q_-, N_- $),  the charges $Q_0, \bar Q_0$ (and $Q_{ax}, Q_{vec}$) (\ref{3.5}) of the $U(1)$ CFT takes the form:
\br
Q_0 &=& {{4\pi n}\o {n+1}}(s+{{j_w}\o n}), \quad s=s_+-s_-, \;\;j_w = j_{\varphi} -2 j_q, 
\quad j_{\varphi} = N_+ - N_{-},\quad j_q =q_+ - q_{-}, \nonu \\
\bar Q_0 &=& -{{4\pi n}\o {n+1}}(\bar s+{{j_{\bar w}}\o n}), \quad 
 \bar s=\bar s_+ - \bar s_-, \quad j_{\bar w} = j_{\varphi} -2 j_{\bar q}, \quad 
 j_{\bar q} =\bar q_+ - \bar q_{-}\nonu \\
 &&s, \bar s \in Z, \quad 
j_{\varphi}, j_q, j_{\bar q}= 0, \pm 1, \cdots \pm (n-1) \;\; {\rm mod} \;\; n
\label{3.36}
\er
The integers $s, \bar s$ denote the winding numbers of $w$ and $ \bar w$, i.e. the number 
of times $w$ winds $S^1$ of radius $r_0 = {{1\o
{2\b_0^2}}}$ when $x$ is running from $-\infty $ to $\infty $. The integers 
$j_w  $ and $j_{\bar w}$ are the $Z_n$-charges of the
vertices $V_w^{s,j_w} = e^{2i \b_0^2 w}$ (or $V_{\bar w}^{\bar s,j_{\bar w}} = e^{2i \b_0^2 \bar w}$).  It is worthwhile 
to mention the relation of $Q_0$ and $\bar Q_0$ with the topological charges of the conformal fields $R^{CFT}= \b_0(w+\bar
w)$ and $\theta^{CFT}_R = {{\b_0}\o 2}(w-\bar w )$, namely:
\br
Q_R^{CFT} = \b_0 \int_{-\infty}^{\infty} \pa_x R^{CFT} dx = {{n+1} \o {4n}}(Q_0 -\bar Q_0), \quad 
Q_{\theta}^{CFT} = \b_0 \int_{-\infty}^{\infty} \pa_x \theta ^{CFT} dx = {{n+1} \o {8n}}(Q_0 +\bar Q_0)
\nonu 
\er

We next remind the vacua structure of the {\it gauged} IM (\ref{1.9})\cite{elek}:
\br
\varphi_l^{(N)} (\pm \infty ) = {{2\pi }\o {\b_0}} {{lN_{\pm}}\o {n}}, \quad \psi_g \chi_g (\pm \infty ) = 0, \quad 
\theta_g^{(L)} (\pm \infty ) = {{\pi L_{\pm}}\o {2\b_0}}.
\nonu
\er
The corresponding $U(1)$ charged topological g-solitons ($\equiv$ gauged solitons ) are characterized by their topological
($j_{\varphi}, j_{\theta}$) and electric $j_{el}$ charges,
\br
Q^{el}& =& {{2n}\o {n+1}}\b_0 \int_{-\infty}^{\infty} \pa_x R_g dx = \b_0^2 j_{el}, \nonu \\
Q_{\theta}^g &= &\b_0 \int_{-\infty}^{\infty} \pa_x \theta_g dx = {{\pi j_{\theta}}\o 2}, \quad j_{\theta} = L_+ - L_-, \nonu
\\
Q_{\varphi_l}^{top}& =& {{2n}\o {\b_0}}\int_{-\infty}^{\infty} \pa_x \varphi_l dx = {{4\pi l}\o {\b_0^2}}j_{\varphi}, 
\quad j_{\varphi} = N_+ - N_- \;\; {\rm mod } \;\; n
\label{3.37}
\er
where $j_{\theta}=0$ for 1-solitons and  $j_{\theta}\neq 0$  for charged breathers \cite{elek}, \cite{tau}. 
 The CFT dressing of such
g-solitons according to eqn. (\ref{3.11}) (with $w$ and $\bar w$ having specific b.c.'s(\ref{3.35}))
 maps each g-soliton to
topological soliton (or string) of the ungauged IM (\ref{1.2}) (called u-soliton), carrying the 
charges of both CFT and
g-solitons:
\br
(j_{el}, j_{\varphi}, j_{\theta} | s, j_w; \bar s, j_{\bar w} )
\nonu
\er 
The one u-solitons are finite energy topological solutions of the IM (\ref{1.2}) that interpolate between two different
ungauged vacua 
\br
(j_{\varphi}|s,j_w; \bar s, j_{\bar w})
\label{3.38}
\er($j_{\theta}=0$ for 1-solitons).  It becomes clear that the $U(1)$-CFT provides each g-vacua $(j_{\varphi}, j_{\theta})$
with an infinite tower of conformal ``states'' (\ref{3.38}) called u-vacua.  The origin of its structure is in the allowed
b.c.'s (\ref{3.30}) for the IM (\ref{1.2}).

\subsection{$U(1)$ CFT solitons}

We are interested in a specific class of topological solutions of $U(1)$ CFT with finite energy (real and positive) and
having eqns. (\ref{3.35}) as b.c.'s for $\Phi, \bar \Phi$ (and $w, \bar w$).  The angular nature of $w_0 =2\b_0^2w$ (and  
$\bar w_0 =2\b_0^2\bar w$), i.e., of $R_0^{CFT} = 2\b_0 R^{CFT}$, is an indication that $w_0, \bar w_0$ we seek  represent
the map of 2-D Minkowski space $M_2$ to the torus $T_2$:
\br
M_2 \rightarrow S_1^{r_0} \otimes  S_1^{r_0}, \quad \quad r_0 = {1\o {2\b_0^2}}
\nonu 
\er
with certain discontinuities (branch cuts) allowed.  Hence they should satisfy 2-D Poisson equation
\br
\pa _{\rho } \pa _{\bar \rho} w_0 = \sum \a_i \d^{(2)} (\rho -\rho_i), 
 \quad \sum \a_i =0
\label{3.39}
\er
 where  $\a_i$ are static charges localized at  $\rho_i$ and 
  $\rho  = e^{a_0z},\quad \bar \rho = e^{a_0\bar z}$ denote the new coordinates. 
  The problem is quite similar to the vortex solutions of 2-D
 Euclidean $U(1)$ CFT \cite{zub} with ``magnetic operators'' $V_w^{s,0} (z_i) = e^{2i \b_0^2 w(z_i)}$ 
 creating
 discontinuity $2\pi s$ at $\rho =\rho_i$.  It turns out that the simplest solution with all the 
 required properties is given by
 the (twisted) Cayley transform:
 \br
 w^{top}=-i\d ln \( {{e^{a_0 z} +i}\o {e^{a_0 z} -i} }\) , \quad \quad
\bar  w^{top}=-i\bar \d ln \( {{e^{a_0 \bar z} +i}\o {e^{a_0 \bar z} -i} }\) 
\label{3.40}
\er
with 
\br
\d = {1\o {\b_0^2}}(s + {{j_w}\o n}), \quad \quad  \bar \d = {1\o {\b_0^2}}(\bar s + {{j_{\bar w}}\o n})
\label{3.41}
\er
and $a_0$ is an arbitrary infrared (IR) scale.  
 In the  pure winding sector ($j_w =0$, i.e., $j_q = {{j_{\varphi}}\o 2}$) we
have 
\br
w_0^{top}(\infty ) =0, \quad \quad  
w_0^{top}(-\infty ) =2\pi s
\nonu 
\er
which confirms the fact that $w_0^{top}$ maps the infinite interval $(-\infty , \infty )$ to a circle.
The energy of the left-moving solitons ($\bar w_0 =0$) takes the form (see eqn. (\ref{3.22})):
\br
E^{CFT}_{L-sol} = \int_{\-\infty}^{\infty} T^{CFT}(z) dx = {{2n \b_0^2}\o {n+1}}\int_{\-\infty}^{\infty} (\pa _x w^{top})^2 dx =
{{4n}\o {(n+1)\b_0^2}} (s + {{j_w}\o n})^2|a_0 |
\label{3.42}
\er
The dimensionless quantity ${{E^{CFT}_{L-sol}}\o {|a_0 |}} = \Delta $ coincides with the conformal dimension of the vortex
operator $V_w^{s, j_w}$ with topological charge 
\br
Q_0 = {{4\pi n}\o {n+1}}(s + {{j_w}\o n})
\label{3.43}
\er
i.e. $\Delta = {{(n+1)} \o {(2\pi \b_0)^2 n}} Q_0^2$.  The general solution of eqn. (\ref{3.39}) (with b.c.'s (\ref{3.35}))
includes also an arbitrary holomorphic (i.e. string oscillators) part $w_{str}$, i.e.,
\br
w = w_{top} + w_{str}, \quad \quad w_{str}(\pm \infty ) =0
\label{3.44}
\er
It has the same charge (\ref{3.43}) as $w_{top}$ but its energy acquires ``string'' contributions from $w_{str}$:
\br
E_{L-string} = {{2n}\o {n+1}} \( {{2}\o {\b_0^2}}(s+ {{j_w}\o n})^2 |a_0| + \b_0^2 \sum_{l\neq 0} Q_l Q_{-l}\)
\label{3.45}
\er
where $\pa w_{str} = \sum_{l\neq 0} Q_l z^l$.

We should mention that $w_{top}$ given by eqn. (\ref{3.40}) is not the most general real nonhomegeneous solution of eqn.
(\ref{3.39}).  One can construct one parameter family $w_{top}^{\a}$ of such solutions with the same b.c.'s (\ref{3.35})
but with arbitrary position of the branch cut:
\br
 w_{top}^{\a}=-i\d_{\a} ln \( {{e^{a_0 z} +e^{i\a}}\o {e^{a_0 z} + e^{-i\a}} }\) , \quad \quad
\d_{\a} = {{(s+{{j_w}\o n})}\o {\b_0^2}}\( {{\pi}\o {2\a}}\) 
\label{3.46}
\er
which for $\a = {{\pi }\o 2}$ concides with (\ref{3.40}).  By construction they carry the same topological charge
(\ref{3.43}), but their energy is $\a$-dependent:
\br
E_{L-sol}^{CFT}(\a ) = {{4n}\o {(n+1)\b_0^2}}(s+{{j_w}\o n} )^2 |a_0 | ({{\pi }\o {2\a}})^2 (1- \a {{\cos (\a )}\o {\sin (\a
)}})
\label{3.47}
\er
and positive for, say,  ${{\pi} \o 4} < \a <\pi $.

The left (and right) solitons are massless by construction, since $E_L=P_L$ (or $E_R =-P_R$ for the right ones).  One could have
however a nontrivial scattering of left and right solitons of rapidities $b_L-b_R \sim 0$, 
\br
E_L = P_L = {1\o 2} M_0e^{b_L}, \quad \quad E_R = -P_R = {1\o 2} M_0e^{-b_R}
\nonu
\er
leading to massive poles at intermediate (crossover) energy scale $\sim M_0^2$ that spoils the infrared  scale (and
conformal) invariance \cite{zz}.
Kinematically such possibility indeed exists, 
\br
M_0^2 = 4 E_L^{CFT} E_R^{CFT}, \quad \quad   b_L-b_R \sim 0
\nonu 
\er
and it has been studied in the context of the marginal  perturbation of $SU(2)$ WZW model in 
ref. \cite{zz}.  We assume that
such phenomenon  ( existence of preferable scale, breaking conformal symmetry at intermediate 
energies) takes place in the IM under
consideration, i.e. together with massless solitons we also have the massive left-right solitons
 (for $\a = {{\pi}\o 2}$)
\br
M_0 = {{8n}\o {(n+1)\b_0^2}}(s+ {{j_w}\o {n}})(\bar s + {{j_{\bar w}}\o n})|a_0 |
\nonu \\
Q_0 = {{4\pi n}\o {n+1}}(s+ {{j_w}\o n}), \quad 
\bar Q_0 = -{{4\pi n}\o {n+1}}(\bar s+ {{j_{\bar w}}\o n})
\label{3.48}
\er
We remind that the CP-invariance (\ref{3.27}) imposes
\br 
s=-\bar s, \quad j_w + j_{\bar w} = 2 (j_{\varphi} - n), \quad j_q + j_{\bar q} = n
\nonu 
\er
The proof of the conjecture of appearence of massive solitons at certain 
 intermediate scale $M_0$ requires the construction of the corresponding left-right S-matrices which is out of the scope
of this paper.

\subsection{Spectrum of the ungauged 1-solitons}

The explicit construction of the u-solitons is based on  eqn. (\ref{3.11}), replacing $w$ and $\bar w$ by the $U(1)$-CFT
solitons (\ref{3.40}) and $\psi_g, \chi_g, R_g = {{n+1}\o {n}}R $  - by the corresponding 1-soliton solutions of the gauged IM
(\ref{1.9}) (see eqns. (\ref{4.5})-(\ref{4.10}) of ref. \cite{elek} ).  The nonconformal GKO formula (\ref{3.20}) allows us to
calculate the energies (and masses) of the ungauged IM 1-solitons in terms of the energies of the constituent gauged and CFT
solitons.  For the left-u-solitons we find 
\br
E_u &=& E_g +E_{L-sol}^{CFT}, \quad P_u = P_g +P_{L-sol}^{CFT}, \quad E_L^{CFT}= P_L^{CFT}, \nonu \\
E_g &=& M_g \cosh b, \quad P_g = - M_g \sinh b, \quad M_u^2 = M_g \( M_g + 2 E_{L-sol}^{CFT} e^{b}\) 
\label{3.49}
\er where $b$ is the g-soliton velocity and $M_g$ its mass \cite{elek}
\br
M_g = {{4\m }\o {\b_0^2}}n | \sin \( {{\b_0^2 j_{el} - 4\pi j_{\varphi}}\o {4n}}\)|
\label{4.50}
\er
Due to the arbitrariness of the $|a_0|$ scale, we can introduce a new scale parameter $m_0$, such that 
$|a_0| = m_0e^{-b}$.  Then the mass $M_u$ of the ungauged 1-soliton takes the form
\br
\( {{M_u}\o {m_0}} \)^2 = \( {{M_g}\o {m_0}} \) \({{M_g}\o {m_0}} + {{8n}\o {(n+1)\b_0^2}}(s+{{j_w}\o n})^2\) 
\label{3.51}
\er
which in fact determines  the dimensionless mass-ratios ${{M_u}\o {m_0}}$ only.  According to eqns. (\ref{3.15}), 
(\ref{3.16}) and (\ref{3.43}) the charge spectrum of the left u-soliton is given by:
\br
Q_{\theta}^u &=& {{n+1}\o {8n}}Q_0 = {{\pi }\o 2}(s+ {{j_w}\o {n}}), \quad Q_{\theta}^g = 0, \nonu \\
Q_R^u &=& {{n+1}\o n}\( \b_0^2 j_{el} + {{2\pi n}\o {n+1}} (s+{{j_w}\o n})\),  \nonu\\
 Q_{el}^u &=& Q_{el}^g = \b_0^2 j_{el}, \quad j_{el} = 0, \pm 1, \pm 2, \cdots
\label{3.52}
\er
Similar formulae take place for the right-u-soliton ($w=0, \bar w \neq 0$).

The spectrum of the left-right u-soliton  combines the energies, masses and charges of the g-soliton
\cite{elek} with the
left-right soliton (\ref{3.48}):
\br
E_u^{L-R} = E_g + E_{L-sol}^{CFT} +E_{R-sol}^{CFT}, \quad M_u^{L-R} = M_g +M_0 \nonu \\
Q_{\theta}^u ={{n+1}\o {8n}}(Q_0 + \bar Q_0), \quad Q_R^u = {{n+1}\o {2n}}\( Q_{el}^u + {1\o 2} (Q_0 -\bar Q_0)\)
\label{3.53}
\er
Together with charged topological u-solitons representing stable strong coupling particles one can also have u-strings
represented by eqns. (\ref{3.11}), but with $w_{top}, \bar w_{top}$ of eqn. (\ref{3.40}) replaced by $w=w_{top} + w_{str}$
(\ref{3.44}).  They have the same charge spectrum as the u-solitons (but with infinite set of charges $Q_k$ and $\bar Q_k$
added) and the energy spectrum including the string oscillator part:
\br
E_u^{string} =E_g + E_{L-string} +  E_{R-string}
\label{3.532}
\er
where $E_{L-string}, ( E_{R-string})$ are given by (\ref{3.45}).

\subsection{$\theta$-terms, dyonic effects and spectral flows}

Similarly to the gauged IM \cite{elek}, one can add to the ungauged IM Lagrangian (\ref{1.2})
 certain topological $\theta$-terms, i.e.,  $
{\cal L}_u^{impr} = {\cal L}_u + \d {\cal L}_u^{top}$ with  $\d {\cal L}_u^{top}$ given by
\br
\d {\cal L}_u^{top}& =& {{\b }\o {8\pi^2}}\( \sum_{k=1}^{n-1} \nu_k^{\varphi} 
\eps^{\mu \nu} \pa_{\mu} \varphi_k \pa_{\nu} ln ( {{\chi_u}\o {\psi_u}}) + 
2\pi \b \sum_{k=1}^{n-1} \tilde \nu_k \eps^{\mu \nu} \pa_{\mu} \varphi_k \pa_{\nu}R_u  \right. \nonu \\
&+& \left.  \pi
\nu^R \eps^{\mu \nu} \pa_{\mu} R_u \pa_{\nu}ln ( {{\chi_u}\o {\psi_u}}) \), 
\label{3.54}
\er
($\nu_k^{\varphi}, \tilde \nu_k, \nu^R$ are real parameters).  They do not change the equations of motion, but contributes to
the charges $Q_{el}, Q_0, \bar Q_0 \rightarrow Q_{el}^{impr},  Q_0^{impr}, \bar Q_0^{impr}$.  For example, the electric
current $I_{\mu}^u$ (\ref{3.6}) (generated by the global $U(1)$ transformations (\ref{3.7})) acquires extra terms, 
\br
I_{\mu}^{u, impr} = I_{\mu}^{u} - {{\b_0^3}\o {4\pi^2}}\eps_{\mu \nu} 
\(\sum_{k=1}^{n-1}\nu_k^{\varphi}  \pa^{\nu}\varphi_k + 2\pi \nu^R \pa^{\nu} R_u \)
\label{3.55}
\er
Then, say for left-u-solitons we have
\br
Q_{el}^{impr} &=& Q_{el} - {{\nu \b_0^2}\o {2\pi}} j_{\varphi} - {{\nu^R} \o {2\pi}}Q_R^u, \quad 
Q_0^{impr} = Q_0 - {{\tilde \nu \b_0^2}\o {\pi}}j_{\varphi} - {{\nu^R \b_0^2}\o {\pi}}Q_{\theta}^u, \nonu \\
Q_R^u &=& {{n+1}\o {2n}}(Q_{el} + {1\o 2}Q_0), \quad Q_{\theta}^u = {{n+1}\o {8n}}Q_0
\label{3.56}
\er
where we have introduced $\nu$ and $\tilde \nu$ as follows
\br
\nu = {{1\o n}}\sum_{k=1}^{n-1} k \nu_k^{\varphi}, \quad \tilde \nu = {{1\o n}}\sum_{k=1}^{n-1} k \tilde \nu_k^{\varphi}
\nonu
\er
Taking into account  eqn. (\ref{3.43}) and that $Q_{el}^{impr} = \b_0^2 j_{el}$ is 
quantized semiclassically \cite{elek}, we
derive the improved charge spectrum
\br
Q_0^{impr} &=& {{4\pi n}\o {n+1}}(1- {{\b_0^2 \nu^R(n+1)}\o {8\pi n}}) (s+ {{j_w}\o n})- {{\b_0^2 \tilde \nu  }\o
{\pi}}j_{\varphi} - {{\b_0^2 \nu^R}\o {2 }}j_{\theta}, \nonu \\
Q_{el} &=& {{\b_0^2 \(j_{el} + {{\nu}\o {2\pi}}j_{\varphi} + {{\nu^R}\o 2}(s+ {{j_w}\o n})\)}\o {1- {{\b_0^2 \nu^R}\o {4\pi
n}}(n+1)}}
\label{3.57}
\er
The interpretation of the parameters $\nu_k^{\varphi}, \tilde \nu_k$ and $\nu^R$ as external constant magnetic fields is
similar to one already presented in refs. \cite{elek}, \cite{new}.

Note that the shift in $Q_0$ by $\tilde \nu j_{\varphi}$ and $\nu_R j_{\theta}$ that  gives $Q_0^{impr}$ (\ref{3.56}) ( and
similar for $\bar Q_0^{impr}$) appears to be the {\it nonconformal} analog of the spectral flow \cite{mo}
\br
\tilde J_3^0 = J_3^0 + {{k\o 4}}\tilde \omega
\label{3.58}
\er
($ \tilde \omega $ is an integer) playing an important role in the description of $AdS_3$-CFT (i.e. the representations of 
$SL(2,R)$ current algebra).

\subsection{Solitons of deformed $SL(2,R)$-WZW model}

As we have mentioned in Sect. 2.3, the particular case $n=1$ and   $\b$ real of the IM (\ref{1.2}) represents integrable
deformation of the $SL(2,R)$-WZW model described by the Lagrangian (\ref{2.29}).  The problem of the vacua structure and
soliton solutions of this IM should be considered separately by the following three reasons:
\begin{itemize}
\item its potential $V_u^{n=1} = m^2 \psi_u \chi_u e^{\b R_u}$ has not distinct zeroes and therefore one cannot expect to
have topological solitons.
\item for real $\b$ the $R^{CFT}$ (and $w, \bar w$) b.c.'s are not periodic  and as a consequence the $U(1)$-CFT solitons
are not topological too.
\item contrary to the $n\geq 2$
IMs the g-solitons of the $n=1$ gauged IM (i.e. Lund-Regge \cite{lund}, \cite{dorey}) 
which are an important ingredient of the u-solitons are
also nontopological.
\end{itemize}
Therefore we have to make certain modifications in the arguments of Sect. 3.3 - 3.5 in order to derive the spectrum of such
nontopological u-solitons.   All the relations between currents, charges and stress-tensors of the gauged and ungauged IM of
Sect. 3.1 and 3.2 are still valid for $n=1$ and $\b$ real.  For example, the field relation (\ref{3.11}) now reads
\br
R_u = R_g + \b (w + \bar w), \quad \theta_u = \theta_g - {{\b }\o 2}(w - \bar w), \nonu \\
\psi_u =\psi_g e^{-{{\b }\o 2}R_g - \b^2 w}, \quad 
\chi_u =\chi_g e^{-{{\b }\o 2}R_g - \b^2 \bar w}
\label{3.59}
\er
Due to the fact that this case ($n=1$, $\b$ real)\footnote{for imaginary $\b =i\b_0$ and $n=1$, i.e., $SU(2,R)$-WZW, however
the following $Z$-transformation $
\psi_u^{\pr} = \psi_u e^{i\pi s}, \quad \chi_u^{\pr} = \chi_u e^{-i\pi s}, \quad R_u^{\pr} = R_u - {{2\pi}\o {\b_0}} s_R $ 
takes place ( $\psi_u^{*} = \chi_u$ in SU(2) case).} we have no analog of the $Z_n$ discrete symmetry (\ref{3.25}), the
b.c.'s of the $R_u$ (and $w, \bar w$) are determined by the global $U(1)_{vector}\otimes U(1)_{axial}$
transformations (\ref{3.28}) and (\ref{3.29}) only, i.e.,
\br
\psi_u^{\pr} = \psi_u e^{-{{\b^2}\o 2} a_R}, \quad 
\chi_u^{\pr} = \chi_u e^{-{{\b^2}\o 2} a_R}, \quad R_u^{\pr} = R_u + \b a_R, \quad \theta_u^{\pr} =  \theta_u, \nonu \\
\psi_u^{\pr \pr } = \psi_u e^{-{{\b^2}} a_{\theta}}, \quad 
\chi_u^{\pr \pr } = \chi_u e^{{{\b^2}} a_{\theta}}, \quad R_u^{\pr \pr } = R_u, \quad \theta_u^{\pr} =  \theta_u + \b
a_{\theta}
\label{3.60}
\er
Therefore the allowed b.c.'s for $w$ and $\bar w$ have the form
\br
w^{\pr} = w + a_{w}, \quad \bar w^{\pr} = \bar w + \bar a_w, \quad a_w + \bar a_w = a_R, \quad 
\bar a_w -  a_w = 2a_{\theta}
\label{3.61}
\er
where $a_w, \bar a_w, a_R, a_{\theta}$ are arbitrary real constants.  As a consequence of (\ref{3.61}) we can
chose
\br
w(\pm \infty ) = a^{\pm}_w, \quad \quad \bar w (\pm \infty ) = \bar a^{\pm }_w
\label{3.62}
\er
which determine the spectrum of the $U(1)$-charges $Q_0, \bar Q_0$, namely
\br
Q_0 = 2\b^2 (a_w^+ - a_w^-) \equiv {{2\pi a}\o {\b^2}}, \quad 
\bar Q_0 = -2\b^2 (\bar a_w^+ - \bar a_w^-) \equiv -{{2\pi \bar a}\o {\b^2}}
\label{3.63}
\er
Since $2 \b R^{CFT}$ and $2\b^2 w$ are {\it not} angular variables in the $SL(2,R)$ case (and the charge spectrum is continuous) the
$w_{top}$ and $\bar w_{top}$ in the form (\ref{3.40}) have no topological meaning.  They map $M_2$ into a rectangle (with
finite area $\pi^2 \d \bar \d $) instead the torus as it is in the case of imaginary $\b =i\b_0$.  Our choice of 
$w_{top}$ and $\bar w_{top}$ is again in the form (\ref{3.40}), but with $\d, \bar \d$-continuous, 
i.e., 
\br
\d = {{2a}\o {\b^2}}, \quad \bar \d = -{{2\bar a}\o {\b^2}}
\label{3.64}
\er
It is dictated by eqns. (\ref{3.39}) and indeed  ensures the required discontinuities.  The energy of such solutions
(``nontopological'' CFT-solitons) is finite and positive:
\br
E_{L-sol}^{CFT} = {{8a^2}\o {\b^2}}|a_0|, \quad E_{R-sol}^{CFT} = {{8\bar a^2}\o {\b^2}}|a_0|
\label{3.65}
\er
According to eqn. (\ref{3.59}) (and similarly to the generic $n\geq 2$ case of Sect. 3.5) the 1-solitons of this deformed
$SL(2,R)$-WZW are a composition of the above ``CFT-solitons'' and the 1-solitons (nontopological) of the gauged $n=1$ IM
(i.e. Lund-Regge model \cite{lund}).  Their semi classical spectrum \cite{dorey} is given by 
\br
M_g = {{4m}\o {\b^2}}\sin ({{\b^2 j_{el}}\o {4}}), \quad \quad Q_{el}^g = \b^2 j_{el}, \quad j_{el} = 0, \pm 1, \pm 2,
\cdots
\label{3.66}
\er
Hence the spectrum of the nontopological left 1-solitons of the IM (\ref{2.29}) has the form
\br
Q_R^u = \b^2 j_{el} + {{a}\o {\b^2}}, \quad Q_{\theta}^u = {{a}\o {2\b^2}}, \quad M_u^2 = 
M_g (M_g + {{16a^2}\o {\b^2}}m_0)
\label{3.67}
\er
The energy of the corresponding $n=1$ string solutions is quite similar to the generic $n$ 
formula (\ref{3.45}),
(\ref{3.532}) with $E_{L-sol}$ replaced by (\ref{3.65}) and $E_g$ - with the Lund-Regge soliton energy.  

It is worthwile to mention  that in the $SU(2)$ case ($\b = i \b_0$, $R_u$ with periodic b.c.'s 
and $\psi_u^{*} = \chi_u$)
the charges $Q_0, \bar Q_0$ are  quantized, i.e. $a=s\in Z$ and the  corresponding CFT-solitons are topological and stable. 
Similarly to the $n\geq 2$ IMs (\ref{1.2}) the deformed $SU(2)$ WZW admits four kinds of 1-soliton solutions:
\begin{itemize}
\item massless topological $U(1)$ CFT solitons and strings
\item massive  solitons of the gauged (Lund-Regge )IM
\item massive composite left (and right ) u-solitons
\item massive left-right u-solitons
\end{itemize}
Whether the composed left-u-solitons of deformed $SL(2,R)$ WZW with spectrum (\ref{3.67}) represent stable strong coupling
particles is an open question.  It is clear however that the corresponding u-solitons of the $SU(2)$ model are indeed
topologicaly stable.

\sect{Multicharged IMs with local and global symmetries}

Among the vast family of dyonic IMs introduced in Sect. 2, we have choosen the simplest $A_n^{(1)}(p=1)$ IM (\ref{1.2})
(with one local $U(1)$ symmetry) in order to demonstrate how the spectrum of its u-solitons can be realized in terms of
the charges, energies , etc of the g-solitons of the gauged IM (\ref{1.9}) and certain $U(1)$ CFT solitons (\ref{3.40}). 
It is natural to address the question of whether the established reduction of the ungauged IM properties to the
corresponding gauged IM combined with $U(1)$ CFT (with specific b.c.'s) takes place for generic multicharged IMs, i.e.,
$\lie_0^0 = U(1)^l$.  Our main attention in the present section is concentrated on the $ \lie_0^0 = U(1)\otimes U(1)$
multicharged IM (\ref{2.13}) and particularly on the intermediate IM (\ref{2.16}) with one local and one global $U(1)$
symmetries.

\subsection{Conserved charges and GKO energy spliting}
 The chiral $U(1)\otimes U(1)$
conserved currents 
\br
J_a = Tr \( g_0^{-1} \pa g_0 \l_a\cdot H^{(0)}\) , \quad 
\bar J_a = Tr \( \bar  \pa g_0 g_0^{-1}\l_a\cdot H^{(0)}\), a=1,n
\nonu
\er
($g_0 \in G_0$ is given by eqn. (\ref{2.12})) of the ungauged IM (\ref{2.13}), generated by the transformations
(\ref{2.14}), (\ref{2.15}), take the following explicit form:
\br
J_1 &=& {{2\b_0}\o {n+1}}(n\pa R_1^u + \pa R_n^u) - i\b_0^2 \psi_1^u \pa \chi_1^u e^{i\b_0(R_1^u-\varphi_1)}, \nonu \\
J_n &=& {{2\b_0}\o {n+1}}(\pa R_1^u + n\pa R_n^u) - i\b_0^2 \psi_n^u \pa \chi_n^u e^{i\b_0(R_n^u-\varphi_{n-2})}, \nonu \\
\bar J_1 &=& {{2\b_0}\o {n+1}}(n\bar \pa R_1^u + \bar \pa R_n^u) - i\b_0^2 \chi_1^u \bar \pa \psi_1^u e^{i\b_0(R_1^u-\varphi_1)}, \nonu \\
\bar J_n &=& {{2\b_0}\o {n+1}}(\pa R_1^u + n\pa R_n^u) - i\b_0^2 \psi_n^u \pa \chi_n^u e^{i\b_0(R_n^u-\varphi_{n-2})}
\label{4.1}
\er
We denote their charges by
\br
Q_{m+1}^a = \oint J_a(z) z^{-m-1} dz, \quad \bar Q_{m+1}^a = \oint \bar J_a(\bar z) \bar z^{-m-1} d\bar z, \quad m \in Z
\label{4.2}
\er
Similarly to the $U(1)$ case (\ref{3.8}), the topological charges
\br
Q_{R_u}^a= \b_0 \int_{-\infty}^{\infty} \pa_x R_a^u dx, \quad a=1,n,
\label{4.3}
\er
the global $U(1)\otimes U(1)$ electric charges $Q_a^{el, u}$:
\br
Q_1^{el, u}&=& i\b_0^2\int_{-\infty}^{\infty}  (\psi_1^u \pa \chi_1^u - \chi_1^u \pa \psi_1^u
)e^{i\b_0(R_1^u-\varphi_1)}dx \nonu \\
Q_n^{el, u}&=& i\b_0^2\int_{-\infty}^{\infty}  (\psi_n^u \pa \chi_n^u - \chi_n^u \pa \psi_n^u
)e^{i\b_0(R_n^u-\varphi_{n-2})}dx
\label{4.4}
\er
and zero modes $Q_0^a, \bar Q_0^a$ of the chiral currents $J^a, \bar J^a$ are related as follows
\br
Q_1^{el, u} = {{2}\o {n+1}}(nQ_{R_u}^{(1)}+ Q_{R_u}^{(n)}) - {1\o 2} (Q_{0}^{(1)}- \bar Q_{0}^{(1)})\nonu \\
Q_n^{el, u} = {{2}\o {n+1}}(Q_{R_u}^{(1)}+ nQ_{R_u}^{(n)}) - {1\o 2} (Q_{0}^{(n)}- \bar Q_{0}^{(n)})
\label{4.5}
\er

The fact that by  construction the gauged IM (\ref{2.19}) is a result of the gauge fixing of the local $U(1)\otimes U(1)$
symmetries (\ref{2.15}) (in the way that $J_a = \bar J_a=0$) and of the consequent elimination of two degrees of freedom
$R_a^u, a=1,n$ lead to the following relation between the gauged   and ungauged   fields
appearing in (\ref{2.19}) and in (\ref{2.13})respectively, 
\br
R_a^u &=& R_a^g + \b_0 (w_a + \bar w_a), \quad \varphi_l^u = \varphi_l^g = \varphi_l, \nonu \\
\psi_a^u &=& \psi_a^g e^{-{1\o 2}i \b_0 R_a^g - i\b_0^2 w_a}, \quad 
\chi_a^u = \chi_a^g e^{-{1\o 2}i \b_0 R_a^g - i\b_0^2 \bar w_a}, \quad a=1,n \nonu \\
\theta^u_a &=& \theta^a_g - {{\b_0}\o 2} (w_a - \bar w_a), \quad 
\theta^u_a = {1\o {2i \b_0}}ln ({{\psi_a^u}\o {\chi_a^u}})
\label{4.6}
\er
The matrix form of these relations is given by eqn. (\ref{2.192}). 
 As in the case of the $U(1)$-models of Sect. 3, eqns.
(\ref{4.6}) are crucial in the construction of the solutions of the
 ungauged IM (\ref{2.13}) in terms of the known solutions
\cite{new} of the gauged IM (\ref{2.19}) and certain $U(1)\otimes U(1)$ -CFT 
solutions (i.e. specific $w_a, \bar w_a$).  An
important byproduct of eqns. (\ref{4.6}) are the following relations between 
the charges (topological and Noether) of both
models (\ref{2.13}) and (\ref{2.19})
\br
Q_{R^u}^{(1)} &=& Q_{R^g}^{(1)} + {{n}\o {4(n-1)}}\( Q_0^{(1)} - \bar Q_0^{(1)}- {{1\o n}}(Q_0^{(n)} - \bar
Q_0^{(n)}) \) \nonu \\
Q_{R^u}^{(n)} &=& Q_{R^g}^{(n)} + {{n}\o {4(n-1)}}\(  Q_0^{(n)} - \bar Q_0^{(n)}- {{1\o n}}(Q_0^{(1)} - \bar
Q_0^{(1)})   \) \nonu \\
Q_{\theta^u}^{(1)} &=& Q_{\theta^g}^{(1)} + {{n}\o {8(n-1)}}\( Q_0^{(1)} + \bar Q_0^{(1)}+ {{1\o n}}(Q_0^{(n)} +\bar
Q_0^{(n)}) \) \nonu \\
Q_{\theta^u}^{(n)} &=& Q_{\theta^g}^{(n)} + {{n}\o {8(n-1)}}\(  Q_0^{(n)} +\bar Q_0^{(n)}+ {{1\o n}}(Q_0^{(1)} + \bar
Q_0^{(1)})   \) 
\label{4.7}
\er
where the charges $Q_0^a, \bar Q_0^a$ have been realized in terms of the asymptotics of the free fields $\Phi^a(z), \bar
\Phi^a (\bar z)$ (i.e., $J^a = \pa \Phi^a, \bar J^a = \bar \pa \bar \Phi^a$):
\br
Q_0^a = 2 \int_{-\infty}^{\infty} \pa _x \Phi^a dx = 2 (\Phi^a (\infty ) - \Phi^a (-\infty )), \nonu \\
\bar Q_0^a = -2 \int_{-\infty}^{\infty} \pa _x \bar \Phi^a dx = -2 (\bar \Phi^a (\infty ) - \bar \Phi^a (-\infty ))
\label{4.8}
\er
which are related to  $w_a$ and $ \bar w_a$ as follows
\br
\Phi^{(1)} &=& {{2\b_0^2}\o {n+1}}(nw_1 +w_n), \quad \bar \Phi^{(1)} = {{2\b_0^2}\o {n+1}}(n\bar w_1 +\bar w_n),
 \nonu \\
\Phi^{(n)} &=& {{2\b_0^2}\o {n+1}}(w_1 +nw_n), \quad \bar \Phi^{(n)} = {{2\b_0^2}\o {n+1}}(\bar w_1 +n\bar w_n)
\label{4.9}
\er
Again, as in the $U(1)$ case (\ref{3.16}), the electric charges $Q_a^{el, u}$ and $Q_a^{el, g}$ of the ungauged and gauged
IMs do coincide:
\br
Q_1^{el, u} = {{2}\o {n+1}}\( nQ_{R^g}^{(1)}+ Q_{R^g}^{(n)}\) = Q_1^{el, g}, \nonu \\
Q_n^{el, u} = {{2}\o {n+1}}\( Q_{R^g}^{(1)}+ nQ_{R^g}^{(n)}\) = Q_n^{el, g}
\label{4.10}
\er
as one can verify by substituting (\ref{4.7}) in (\ref{4.5}).

The generalization of the nonconformal GKO formulae (\ref{3.20}) and (\ref{3.21})
to the case of multicharged IMs (\ref{2.13}) and (\ref{2.19}) is straightforward.  Substituting eqns. (\ref{4.6}) in the
ungauged IM stress-tensor:
\br
T_{p=2}^{u} &=& {1\o 2} \eta_{ij} \pa \varphi_i \pa \varphi_j + \pa \psi_1^u \pa \chi_1^u e^{\b (R_1^u - \varphi_1)}+ 
\pa \psi_n^u \pa \chi_n^ue^{\b (R_n^u - \varphi_{n-2})} \nonu \\
&+& {{n}\o {2(n+1)}} \( (\pa R_1^u)^2 + (\pa R_n^u)^2
+ {2\o n} \pa R_1^u\pa R_n^u \) + V_u 
\label{4.11}
\er
and $\bar T_{p=2}^{u} = T_{p=2}^{u} (\pa \rightarrow \bar \pa )$ and taking into account the 
constraints \cite{new}
\br
J(z)= Tr ((g_0^f)^{-1} \pa g_0^f \l_a \cdot H^{(0)}) = 
\bar J(z)= Tr (\bar \pa g_0^f (g_0^f)^{-1}\l_a \cdot H^{(0)}) = 0, \nonu
\er
we realize that the following GKO-spliting 
\br
T_{p=2}^{u} = T_{p=2}^{g} + \b_0^2 {{n}\o {2(n+1)}} \( (\pa w_1)^2 + (\pa w_n)^2 + {2\o n}(\pa w_1)(\pa w_n)\) 
\label{4.12}
\er
 takes place.  We have denoted by $T_{p=2}^{g}$ the canonical stress-tensor of the gauged
IM derived from its Lagrangian (\ref{2.19}), 
\br
T_{p=2}^{g} &=& {1\o 2} \eta_{ij} \pa \varphi_i \pa \varphi_j + 
{1\o {\Delta}}\( \pa \psi_1^g \pa \chi_1^g e^{-\b \varphi_1} (1+ \b_0^2 {{n}\o {2(n-1)}} \psi_n^g \chi_n^g e^{-\b
\varphi_{n-2}}) \right. \nonu \\
&+& \left.
\pa \psi_n^g \pa \chi_n^g e^{-\b \varphi_{n-2}} (1+ \b_0^2 {{n}\o {2(n-1)}} \psi_1^g \chi_1^g e^{-\b
\varphi_{1}})\right. \nonu \\
&+& \left. \b^2 {{1}\o {2(n-1)}}  (\chi_1^g \psi_n^g \pa \psi_1^g \pa \chi_n^g + 
\chi_n^g \psi_1^g \pa \psi_n^g \pa \chi_1^g)e^{-\b (\varphi_1 + \varphi_{n+2})}\) + V_g
\label{4.13}
\er
One can further diagonalize the $U(1)\otimes U(1)$-CFT stress-tensor
\br
T_{p=2}^{CFT} = \b_0^2 {{n}\o {2(n+1)}} \( (\pa w_1)^2 +(\pa w_n)^2 + {{2}\o n}\pa w_1 \pa w_n \) , 
\quad \bar \pa T_{p=2}^{CFT}=0
\nonu
\er
by introducing new fields $w_{\pm}$:
\br
w_{\pm} = {1\o 2} (w_1 \pm w_n ), \quad \quad \bar w_{\pm} = {1\o 2} (\bar w_1 \pm \bar w_n )
\label{4.14}
\er
The result is:
\br
T_{p=2}^{CFT} = \b_0^2  \( (\pa w_+)^2 +{{n-1}\o {n+1}}(\pa w_-)^2 \)  = T_{+}^{CFT} + T_{-}^{CFT}, 
\label{4.15}
\er
and the same for $\bar T_{p=2}^{CFT} = T_{p=2}^{CFT} (\pa \rightarrow \bar \pa )$.

The intermediate IM (\ref{2.16}), introduced in Sect. 2, is a 
result of the gauge fixing of one of the local $U(1)$ symmetries, say,
\br
J_+ &=& {1\o 2} Tr ((g_0^{int})^{-1}\pa g_0^{int} (\l_1 + \l_n )\cdot H^{(0)} ) = 0, \nonu \\
\bar J_+ &=& {1\o 2} Tr (\bar \pa g_0^{int} (g_0^{int})^{-1}(\l_1 + \l_n )\cdot H^{(0)} ) = 0
\label{4.16}
\er
of the ungauged IM (\ref{2.13}).  Its Lagrangian is invariant under chiral $U(1)$ (spanned by 
${1\o 2}(\l_1 - \l_n )\cdot H^{(0)}$) transformation  (\ref{2.18})  with conserved  current $\bar \pa J_- = \pa
\bar J_- =0$
\br
J_- &=& {1\o 2} Tr \( (g_0^{int})^{-1}\pa g_0^{int} (\l_1 - \l_n )\cdot H^{(0)} \)  =
2 \b_0 ({{n-1}\o {n+1}}) \Delta_0 \pa \bar R_u \nonu \\
&-& i \b_0^2 \( (1 + {{\b^2}\o 2}\bar \psi_n \bar \chi_n e^{-i\b_0 (\bar R + \varphi_{n-2})} )
\bar \psi_1 \pa \bar \chi_1 e^{i\b_0 (\bar R- \varphi_1)} \right. \nonu \\
&-& \left. 
(1 + {{\b^2}\o 2}\bar \psi_1 \bar \chi_1 e^{-i\b_0 ( \varphi_{1} - \bar R)} )
\bar \psi_n \pa \bar \chi_n e^{-i\b_0 (\bar R+ \varphi_{n-2})}\)\nonu \\
  \bar J_- &=& J_- (\pa \rightarrow \bar \pa, \psi_a \rightarrow \chi_a ) = 
{1\o 2} Tr \( \bar \pa g_0^{int}(g_0^{int})^{-1} (\l_1 - \l_n )\cdot H^{(0)} \) 
\label{4.17}
\er
where
\br
J_- =\pa \Phi_-, \quad \bar J = \pa \Phi_-,  \quad 
\Phi_- = 2\b_0^2 {{n-1}\o {n+1}} w_-, \quad \bar \Phi_- =2\b_0^2 {{n-1}\o {n+1}} \bar w_-. 
\nonu
\er
They are also invariant under global $U(1)$ transformation 
\br
\bar \psi^{\pr}_a = \bar \psi_a e^{i\b_0^2 \eps_+}, \quad \bar \chi^{\pr}_a = \bar \chi_a e^{-i\b_0^2 \eps_+}, \quad 
\bar R^{\pr}_u = \bar R_u, \quad \varphi_l^{\pr} = \varphi_l
\label{4.18}
\er
We denote the charges of the $J_-$ and $ \bar J_-$ currents by
\br
Q_{m+1}^- = \oint J_-(z)z^{-m-1} dz, \quad \bar Q_{m+1}^- = 
\oint \bar J_-(\bar z) \bar z^{-m-1} d\bar z, \quad  \nonu 
\er
i.e., we have for their zero modes:
\br
Q_0^- = 2 \int_{-\infty}^{\infty} \pa_x \Phi_- dx, \quad \bar Q_0^- = -2 \int_{-\infty}^{\infty} \pa_x \bar \Phi_- dx
\label{4.19}
\er
The global $U(1)$-charges are now given by 
\br
Q_{\pm}^{el, u} = {1\o 2} (Q_1^{el, u} \pm Q_n^{el, u})
\label{4.20}
\er
Taking into account eqn. (\ref{4.10}), we derive the following relation 
\br
Q_+^{el, u} &=& 2 Q_{\tilde R_g} = Q_+^{el, g}, \quad 
Q_-^{el, u} =2 {{n-1}\o {n+1}}Q_{\bar R_g} = Q_-^{el, g}
\label{4.202}
\er
between $Q_{\pm}^{el, u}$  and the topological charges
\br
Q_{\tilde R_g} = \b_0 \int_{-\infty}^{\infty} \pa_x \tilde R_g dx, \quad \tilde R_g = {1\o 2}( R_1^g + R_2^g), \nonu \\
Q_{\bar R_g} = \b_0 \int_{-\infty}^{\infty} \pa_x \bar R_g dx, \quad \bar R_g = {1\o 2}( R_1^g - R_2^g)
\label{4.2022}
\er
Note that the topological charge $Q_{\bar R_u}$, i.e., 
\br
Q_{\bar R_u}= \b_0 \int_{-\infty}^{\infty} \pa_x \bar R_u dx 
\nonu 
\er
is not proportional to the electric charge $Q_-^{el, u}$, due to the more general relation
\br
{1\o 2}(Q_0^- - \bar Q_0^-) = 2 ({{n-1}\o {n+1}})Q_{\bar R_u} - Q_-^{el, u}
\label{4.21}
\er
which follows from eqn. (\ref{4.17}).

According to eqn. (\ref{2.172}) the fields of the intermediate 
model $\bar \psi_a, \bar \chi_a, \bar R$ can be  realized in terms of the
fields of the ungauged IM $ \psi_a^u,  \chi_a^u,  R^u_a$ and the free fields $w_+$ and $ \bar w_+$:
\br
\tilde R_u &=& \tilde R_g^{int} + \b_0 ( w_+ + \bar w_+), \quad \varphi_l^u = \varphi_l^{int}, \quad 
\bar R_u = \bar R_u^{int} = \bar R, \nonu \\
\psi_a^u &=& \bar \psi_a e^{-i \b_0^2 w_+ - {{i}\o 2} \b_0\tilde R_g}, \quad 
\chi_a^u = \bar \chi_a e^{-i \b_0^2 \bar w_+ - {{i}\o 2}\b_0 \tilde R_g}
\label{4.22}
\er
Similarly we can relate the intermediate IM fields with  those of the gauged IM (\ref{1.9})
 $\psi_a^g,  \chi_a^g,  R^g_a$, taking into account eqns. (\ref{4.22}), (\ref{4.6}) and (\ref{4.14}):
 \br
 \bar R&=& \bar R_g + \b_0(w_- + \bar w_-), \quad \varphi_l^{int} = \varphi_l^g, \quad  
  \bar \psi_1=\psi_1^g e^{-i \b_0^2 w_- - {{i}\o {2}}\b_0\bar R_g},\nonu \\
  \bar \psi_n&=&\psi_n^g e^{i \b_0^2 w_- + {{i}\o {2}}\b_0\bar R_g},\quad
\bar \chi_1=\chi_1^g e^{-i \b_0^2 w_- - {{i}\o {2}}\b_0 \bar R_g},\nonu \\
  \bar \chi_n&=&\chi_n^g e^{i \b_0^2 w_- + {{i}\o {2}}\b_0 \bar R_g},\quad
  \bar \theta_a = \theta_a^g \mp {{\b_0}\o 2}(w_- - \bar w_-), \quad a=1,n
\label{4.23}
\er
Therefore the topological charges of the intermediate IM can be realized in terms of the corresponding gauged IM charges and
the $J_-$ and $ \bar J_-$ zero modes:
\br
Q_{\bar R_u} = Q_{\bar R_g} + {{n+1}\o {4(n-1)}} (Q_0^- - \bar Q_0^-), \nonu \\
Q_{\theta_u}^{(1)} = Q_{\theta_g}^{(1)} -{{n+1}\o {8(n-1)}} (Q_0^- + \bar Q_0^-), \nonu \\
Q_{\theta_u}^{(n)} = Q_{\theta_g}^{(n)} +{{n+1}\o {8(n-1)}} (Q_0^- + \bar Q_0^-)
\label{4.24}
\er
Finally, the GKO energy spliting for the intermediate IM takes the form
\br
T^u_{p=2} =T^{int}_{p=2} + \b_0^2 (\pa w_+)^2, \quad 
T^{int}_{p=2} = T^g_{p=2} + \b_0^2({{n-1}\o {n+1}}) (\pa w_-)^2
\label{4.25}
\er
The stress-tensor of the intermediate IM (\ref{2.16}) is given by 
\br
T^{int}_{p=2} &=& {1\o 2}\eta_{ij} \pa \varphi_i \pa \varphi_j + ({{n-1}\o {n+1}})(\pa \bar R)^2 + {{1}\o {\Delta_0}}
\( (1+ {{\b_0^2}\o {4}}\bar \psi_n \bar \chi_n e^{-i\b_o (\bar R +\varphi_{n-2})})e^{i\b_0 (\bar R- \varphi_1)}\pa \bar
\psi_1 \pa \bar \chi_1 \right. \nonu \\
&+& \left.  (1+ {{\b_0^2}\o {4}}\bar \psi_1 \bar \chi_1 e^{-i\b_o (\bar R -\varphi_{1})})e^{i\b_0 
(\bar R+ \varphi_{n-2})}\pa \bar
\psi_n \pa \bar \chi_n \right. \nonu \\
&-& \left.  {{\b_0^2}\o 4} e^{-i\b_0 (\varphi_1 + \varphi_{n-2})} 
(\bar \psi_1 \bar \chi_n \pa \bar \chi_1 \pa \bar \psi_n + 
\bar \psi_n \bar \chi_1 \pa \bar \chi_n \pa \bar \psi_1)\) + V_{int}
\label{4.26}
\er
and $T^u_{p=2}$ and $T^g_{p=2}$ by eqns. (\ref{4.11}) and (\ref{4.13}) respectively.

\subsection{Vacua Structure and 1-solitons of the $U(1)\otimes U(1)$ ungauged IM}

Together with the local (and global) $U(1)\otimes U(1)$ symmetries (described in Sect. 4.1)
 the ungauged IM Lagrangian (\ref{2.13}) is also invariant under discrete $Z_2\otimes Z_2\otimes Z_{n-1}$ transformations (for imaginary
 coupling $\b =i\b_0$).  The action of the $Z_{n-1}$ group on the fields $\varphi_l, \psi_a^u, \chi_a^u$ and $R_a^u$ is quite similar to the
 $Z_n$ transformations (\ref{3.25}) for the $U(1)$ model (i.e. $p=1$) (\ref{1.2}).  It is easy to check that ${\cal L}_{p=2}^u$ (\ref{2.13})
 remains invariant under the following $Z_{n-1}$ transformation:
 \br
 \varphi_l^{\pr} &=& \varphi_l + {{2\pi}\o {\b_0}}{{lN}\o {n-1}}, \quad l=1,2, \cdots n-2, \nonu \\
 (R_1^u)^{\pr}&=& R_1^u + {{2\pi }\o {\b_0}}(s_R^{(1)} + {{N-q_1 -\bar q_1}\o {r-1}}), \nonu \\
 (R_n^u)^{\pr}&=& R_n^u + {{2\pi }\o {\b_0}}(s_R^{(n)} - {{N-q_n -\bar q_n}\o {r-1}}), \nonu \\
(\psi_1^u)^{\pr}&=& e^{\pi i ({{2q_1}\o {n-1}} + s_1)}\psi_1^u, \quad \quad 
(\chi_1^u)^{\pr}=e^{\pi i ({{2\bar q_1}\o {n-1}} + \tilde s_1)}\chi_1^u, \nonu \\
(\psi_n^u)^{\pr}&=& e^{-\pi i ({{2q_n}\o {n-1}} + s_n)}\psi_n^u, \quad \quad 
(\chi_n^u)^{\pr}= e^{-\pi i ({{2\bar q_n}\o {n-1}} + \tilde s_n)}\chi_n^u
\label{4.27}
\er
where $s_R^{(a)}, s_a, \tilde s_a$, ($s_a \pm \tilde s_a = 2s^{(a)}_{\pm})$ are integers and 
$q_a, \bar q_a, N=0, 1, \cdots n-2 \;\; {\rm mod}\;\; (n-1)$ are the $Z_{n-1}$ charges of the fields $\psi_a^u, \chi_a^u$ and
$e^{i\b_0\varphi_1}$ respectively.  The first $Z_2$ acts as field reflections:
\br
\varphi_l^{\pr \pr} &=& \varphi_{n-l-1}, \quad (R_1^u)^{\pr \pr } = R_n^u, \quad (R_n^u)^{\pr \pr } = R_1^u, \nonu \\
(\psi_1^u)^{\pr \pr } &=& \psi_n^u, \quad (\psi_n^u)^{\pr \pr } = \psi_1^u, \quad 
(\chi_1^u)^{\pr \pr } = \chi_n^u, \quad (\chi_n^u)^{\pr \pr } = \chi_1^u,
\label{4.28}
\er
i.e. interchanging the $Z_{n-1}$  charges of $\varphi_1$ and $\varphi_{n-2}$, $\psi_1^u$ and $\psi_n^u$, $R_1^u$ and $R_n^u$, etc.
\br
N \rightarrow n-1-N, \quad q_a \rightarrow n-1-q_a, \quad \bar q_a \rightarrow n-1-\bar q_a
\nonu
\er
This extends the $Z_{n-1}$ to the 
 diedral group $D_{n-1}$.  The other $Z_2$ group represents the following  CP-transformations:
\br
\varphi_l^{\pr \pr \pr } &=& \varphi_{l}, \quad (R_a^u)^{\pr \pr \pr} = R_a^u,\quad 
(\psi_a^u)^{\pr \pr \pr} = \chi_a^u, \quad (\chi_a^u)^{\pr \pr \pr} = \psi_a^u,\nonu \\
Px &=& -x, \quad P\pa = \bar \pa, \quad P^2 =1
\label{4.29}
\er

The vacuum solutions for the ungauged IM (\ref{2.13}) are given by 
\br
\varphi_l^{(N)} = ({{2\pi }\o {\b_0}}) {{lN}\o {n-1}}, \quad \psi_a \chi_a = 0, \quad R_a^u = \b_0 a_R^a, \quad \theta_a^u = \b_0 a_{\theta}^a
\label{4.30}
\er
For such field configuration the potential $V^u_{p=2}$ (\ref{2.132}) shows $(n-1)$-distinct zeroes.  By chiral   $U(1)\otimes U(1)$
transformations (\ref{2.15})  one map these constant vacua solutions into a special class of conformal  invariant solutions 
\br
\varphi_l^{(N)} &=& {{2\pi }\o {\b_0}}{{lN}\o {n-1}}, \quad \psi_a^u \chi_a^u = 0 \nonu \\
R_{a}^{u,CFT} &=& \b_0 (w_a + \bar w_a), \quad \theta_a^{u, CFT} = {{\b_0}\o 2} (w_a - \bar w_a), \quad a=1,n
\label{4.31}
\er
representing string (2-d free fields $w_a, \bar w_a$) in flat 2-d target  space, i.e., $U(1)\otimes U(1)$-CFT.  Among all possible
string-like and particle-like finite energy ($E^{CFT} = \pm P^{CFT}$) solutions of this CFT, we seek for 
a special family of charged 
topological  massless solitons.  As in the case of $U(1)$-CFT (see Sect. 3.3 and 3.4) crucial  for the existence of such solutions are the
nontrivial b.c. for $w_a, \bar w_a$, supported by certain discrete symmetries ($Z_{n-1}$ in our case).  It is  not necessary to impose an
appropriate b.c., since they are already encoded in eqns. (\ref{4.6}) and (\ref{4.27}).  Taking into account the $Z_{n-1}$ transformations of the
gauged IM (\ref{2.13}) fields \cite{new}:
\br
(\psi_1^g)^{\pr}&=& e^{\pi i ({{N}\o {n-1}} + s_1^g)}\psi_1^g, \quad \quad 
(\chi_1^g)^{\pr}=e^{\pi i ({{N}\o {n-1}} + \tilde s_1^g)}\chi_1^g, \nonu \\
(\psi_n^g)^{\pr}&=& e^{-\pi i ({{N}\o {n-1}} + s_n^g)}\psi_n^g, \quad \quad 
(\chi_n^g)^{\pr}= e^{-\pi i ({{N}\o {n-1}} + \tilde s_n^g)}\chi_n^g \nonu \\
(R_a^g)^{\pr}&=&R_a^g, \quad \varphi_l^{\pr} = \varphi_l
\label{4.32}
\er
we derive the following discrete transformations  for $w_a, \bar w_a$,
\br
w_a^{\pr} =w_a + {{\pi}\o {\b_0^2}}(s_a + \eps_a {{N-2q_a}\o {n-1}}), \quad 
\bar w_a^{\pr} =\bar w_a + {{\pi}\o {\b_0^2}}(\bar s_a + \eps_a {{N-2\bar q_a}\o {n-1}}), 
\label{4.33}
\er
where $\eps_1 = -\eps_n=1$.  The $Z_{n-1}$ properties (\ref{4.33}) of $w_a, \bar w_a$ (and of the diagonal fields (\ref{4.14}) $w^{\pm}$ and $\bar
w^{\pm}$) allow us to distiguish an important class of b.c. for the $U(1)\otimes U(1)$-CFTs free fields $\Phi^a(z)$ and 
$\bar \Phi^a(\bar z)$ of eqns. (\ref{4.9}):
\br
\Phi_+(\pm \infty ) &=& \pi  (s_1^{\pm} + s_n^{\pm} + {{2(q_n^{\pm} - q_1^{\pm})}\o {n-1}}), \nonu \\
\Phi_-(\pm \infty ) &=& \pi {{n-1}\o {n+1}} (s_1^{\pm} - s_n^{\pm} + {{2(N_{\pm} -q_1^{\pm} - q_n^{\pm})}\o {n-1}})
\label{4.34}
\er
where $\Phi_{\pm}$ are defined as follows
\br
\Phi_+ &=& {1\o 2}(\Phi^{(1)} + \Phi^{(n)}) = 2\b_0^2 w_+, \nonu \\
\Phi_- &=& {1\o 2}(\Phi^{(1)} - \Phi^{(n)}) = 2{{n-1}\o {n+1}}\b_0^2 w_- ,\nonu \\
J_{\pm} &=&\pa \Phi_{\pm}, \quad J_{\pm} = {1\o 2}(J_1 \pm J_n)
\label{4.35}
\er
Therefore the CFT solutions interpolating between two vacua 
\br
\( \Phi_+(\infty ), \Phi_-(\infty ) \)  \rightarrow \( \Phi_+(-\infty ), \Phi_-(-\infty ) \)
\nonu
\er
carry nontrivial topological charges $Q_0^{\pm}$ and $\bar Q_0^{\pm}$
\br
Q_0^+ &=& 2\int_{-\infty}^{\infty} \pa_x \Phi_+dx = 4\pi (s_+ + {{j_w^-}\o {n-1}}), \nonu \\
Q_0^- &=& 2\int_{-\infty}^{\infty} \pa_x \Phi_-dx = 4{{n-1}\o {n+1}}\pi (s_- + {{j_w^+}\o {n-1}}), \nonu \\
 \bar Q_0^+ &=& -4\pi (\bar s_+ + {{j_{\bar w}^-}\o {n-1}}), \quad \bar Q_0^- =-4\pi {{n-1}\o {n+1}}(\bar s_- 
 + {{j_{\bar w}^+}\o {n-1}})
\label{4.36}
\er
where $s_1\pm s_n = 2s_{\pm},\; \; 2j_{w}^{\pm} = j_{w}^{(1)} \pm j_{w}^{(n)}, \; \;   j_{w}^{a } = j_{\varphi} -2 j_q^{a}, \;\; j_{\varphi}
=N_+-N_-, \;\; j_q^a =q_+^a - q_-^a$.
It becomes clear that the conformal vacua of the $U(1)\otimes U(1)$-CFT are characterized by the winding numbers $s_{\pm}$ and $\bar s_{\pm}$
 of the fields $2\b_0^2 w_{\pm}$ and $2\b_0^2 \bar w_{\pm}$ and by the $Z_{n-1}$ charges $j_w^{\pm}, \; j_{\bar w}^{\pm}$ of the vertices
 $V_{\pm}^{s_{\pm}, j_{w}^{\pm}} = e^{2i \b_0^2 w_{\pm}}$ and  $\bar V_{\pm}^{\bar s_{\pm}, j_{\bar w}^{\pm}} = e^{2i \b_0^2 \bar w_{\pm}}$.
Similarly to the $U(1)$-case of Sect. 3.4, the topological solitons of the   $U(1)\otimes U(1)$-CFT with all the above
properties are nothing but the map of 2-d dimension space $M_2$ to 4-torus $T_4 = (S_1^{r_0})^4, r_0  ={{1 }\o {2\b_0^2}}$.
Their explicit form is a generalization of the 1-soliton $w_{top}$ (\ref{3.40}) of $U(1)$-CFT:
\br
w_{\pm}^{top} &=& -i \d_{\pm}ln \( {{{e^{a_0z}+i}}\o {{e^{a_0z}-i}}}\), \quad  
\bar w_{\pm}^{top} = -i \bar \d_{\pm}ln \( {{{e^{a_0\bar z}+i}}\o {{e^{a_0\bar z}-i}}}\), \nonu \\
\d_{\pm} &=& {{1}\o {\b_0^2}} (s_{\pm} + {{j^{\mp}_w}\o {n-1}}), \quad 
\bar \d_{\pm} = {{1}\o {\b_0^2}} (\bar s_{\pm} + {{j^{\mp}_{\bar w}}\o {n-1}})
\label{4.37}
\er
We next calculate the energy $E_{L-sol}^{p=2}$ of the left moving massless solitons (see eqn. (\ref{4.15})):
\br
E_{L-sol}^{p=2}&=& \int_{-\infty}^{\infty} T_{P=2}^{CFT} dx = 4\b_0^2 \int_{-\infty}^{\infty} \( (\pa_x w_+)^2 + {{n-1}\o
{n+1}}(\pa_x w_-)^2 \) dx \nonu \\
&=& 8\b_0^2 |a_0| ( \d_+^2 + {{n-1}\o {n+1}} \d_-^2 )
\nonu
\er
or explicitly:
\br
{{E_{L-sol}^{p=2}}\o {|a_0|}}&=& {{8}\o {\b_0^2}} \( (s_+ + {{j_w^-}\o {n-1}})^2 + 
{{n-1}\o {n+1}}(s_- + {{j_w^+}\o {n-1}})^2\) 
\label{4.38}
\er
One can also have finite energy string-like solutions with the same topological charges (\ref{4.36}), but admiting arbitrary
holomorphic parts $w_{str}^{\pm}$ of $w^{\pm}$, i.e.,
\br
w^{\pm} = w_{\pm}^{top} + w_{str}^{\pm}, \quad w_{str}^{\pm}(\pm \infty )=0
\nonu
\er
The energy of such string solutions gets contributions from $\pa w_{str}^{\pm} = \sum_{k\neq 0} Q_k^{\pm} z^k $:
\br
E_{L-string}^{p=2} =E_{L-sol}^{p=2} + \b_0^2 \sum_{k\neq 0} \( Q_k^+Q_{-k}^+ + {{n-1}\o {n+1}}Q_k^-Q_{-k}^-\)
\label{4.382}
\er
We should mention that the most general particle-like 1-soliton is represented by the following two parameter family of
solutions:
\br
w_{top}^{\pm} (\a_{\pm} ) =-i \d_{\pm}(\a_{\pm}) ln \( {{e^{a_0z}+ e^{i\a_{\pm}}}\o {e^{a_0z}+ e^{-i\a_{\pm}}}}\), \quad 
\d_{\pm}(\a_{\pm}) = {{\pi }\o {2\a_{\pm}}}\d_{\pm}
\label{4.39}
\er
They have the same topological charges (\ref{4.36}) as $w_{top}^{\pm} (\a_{\pm}=\pi /2) = w_{top}^{\pm}$, but different
energies (for $\a_{\pm} \neq \pi /2, \; \pi /4 < \a_{\pm} <\pi $)
\br
E_{L-sol}^{p=2}(\a_+, \a_-) &=& {{2|a_0| \pi^2}\o {\b_0^2}} \( {{1}\o {\a_+^2}} (s_+ + {{j_w^-}\o {n-1}})^2 \( 1-\a_+cotg (\a_+
\) ) \right. \nonu \\
&+& \left. {{1}\o {\a_-^2}}\( {{n-1}\o {n+1}}\) (s_- + {{j_w^+}\o {n-1}})^2\( 1-\a_- cotg (\a_- \) )\)
\label{4.40}
\er
which  coincides with (\ref{4.38}) for $\a_{\pm} = \pi /2$.  

Given the 1-solitons \cite{new} of the gauged IM (\ref{2.19}) and the $U(1)\otimes U(1)$-CFT topological massless solitons
(\ref{4.37}) and (\ref{4.39}) we can  construct the ungauged 1-solitons of the IM (\ref{2.13})  according to eqns.
(\ref{4.6}).  Then the GKO formula (\ref{4.12}) allows us to calculate the energy of say, left-u-solitons (i.e.,
$w^{\pm}\neq 0, \bar w^{\pm} =0$):
\br
E_u ^{p=2} &=& E_g^{p=2} + E_{L-sol}^{CFT}(p=2),\nonu \\
P_u^{p=2} &=& P_g^{p=2} + P_{L-sol}^{CFT}(p=2), \quad \quad E_L^{CFT}=P_L^{CFT}, \nonu \\
E_g ^{p=2} &=& M_g^{p=2} \cosh (b), \quad P_g^{p=2} = -M_g^{p=2} \sinh (b)
\label{4.402}
\er
where $b$ is the velocity of the gauged IM 1-solitons and $M_g$ their mass \cite{new}:
\br
M_g^{p=2} = {{4m(n-1)}\o {\b_0^2}}|\sin {{4\pi j_{\varphi} - \b_0^2 j_{el}}\o {4(n-1)}}|
\nonu 
\er
For $|a_0|= m_0e^{-b}$, the mass of u-solitons is given by 
\br
\({{M_u^{p=2}}\o {m_0}}\)^2 ={{M_g^{p=2}}\o {m_0}}\( {{M_g^{p=2}}\o {m_0}} + 2 {{E_{L-sol}^{CFT}(p=2)}\o {m_0}}\)
\label{4.41}
\er
Taking into account the charge relations (\ref{4.5}), (\ref{4.7}) and (\ref{4.10}) we have established in Sect. 4.1, we
derive the charge spectrum of the left-u-soliton:
\br
Q_0^+ &=& 4\pi (s_+ + {{j_w^-}\o {n-1}}), \quad Q_0^- = 4\pi {{n-1}\o {n+1}}(s_- + {{j_w^+}\o {n-1}}), \nonu \\
Q_{el}^- &=& {1\o 2} (Q_1^{el}- Q_n^{el}) = {{\b_0^2}\o 2}j_{el}, \quad
 Q_{el}^+ = {1\o 2} (Q_1^{el}+ Q_n^{el}) = {{\b_0^2}\o 2}\d_0, \nonu \\
 Q_{R_u}^- &=& {1\o 2} (Q_{R_u}^{(1)}- Q_{R_u}^{(n)}) = {{n+1}\o {4(n-1)}}(\b_0^2 j_{el} + Q_0^-), \nonu \\
 Q_{R_u}^+ &=& {1\o 2} (Q_{R_u}^{(1)}+ Q_{R_u}^{(n)}) = {1\o 4}(2\b_0^2 \d_0 + Q_0^+), \nonu \\
 Q_{\theta_u}^+ &=& {{n+1}\o {4(n-1)}} Q_0^+, \quad Q_{\theta_u}^- = {{1}\o {4}} Q_0^-
\label{4.42}
\er
where $Q_{el}^{\pm}$ (i.e. $j_{el} \in Z$ and $\d_0 \in R$) represent the semiclassical electric charges of the gauged
1-solitons \cite{new}.

\subsection{Soliton spectrum of the intermediate IM}

The intermediate model (\ref{2.16}) arises from the ungauged IM (\ref{2.13}) by axial gauge fixing one of the chiral $U(1)$
symmetries, namely by imposing the constraints  $J_+ = \bar J_+ =0$ (see eqn. (\ref{4.16})) and keeping 
$J_-$ and $ \bar J_-$
unconstrained.  The same way we have constructed u-solitons as conformal dressing by $w_{top}^a$ (\ref{4.37}) of the
1-solitons \cite{new} of the gauged IM (\ref{2.19}), the intermediate 1-solitons can be constructed by $U(1)$ CFT-dressing of
the same g-solitons by $w_{top}^-$ and $\bar w_{top}^-$ according to eqns. (\ref{4.23}).  The relations between the solitons
of the gauged, intermediate and ungauged IMs, established in Sect. 4.1 provide an alternative construction  of these int-solitons 
as a reduction
of the u-solitons by requiring that $w^+ = \bar w^+ =0$, i.e.,
\br
q_1=q_n=q, \quad  \bar q_1=\bar q_n=\bar q, \quad s_+ = \bar s_- = 0
\nonu 
\er
In order to make transparent the properties of these solitons (and the origin of the b.c.'s for $w^-, \bar w^-$) {we shall
derive the vacua structure of the IM (\ref{2.16}) and its discrete symmetries independently of the relations with the
ungauged IM (\ref{2.13}) }.  The $Z_{n-1}$ transformations that leave invariant the Lagrangian (\ref{2.16}) have the form:
\br
\bar \psi_a^{\pr} &=& \bar \psi_a e^{i\pi \eps_a ({{2q}\o {n-1}}+ s_a)}, \quad 
\bar \chi_a^{\pr} = \bar \chi_a e^{i\pi \eps_a ({{2\bar q}\o {n-1}}+ \bar s_a)}, \quad \eps_1 = -\eps_n =1, \quad s_a \pm \bar
s_a = 2 L_a, \nonu \\
\bar R^{\pr} &=& \bar R + {{2\pi }\o {\b_0}} (s_R + {{N-q -\bar q}\o {n-1}}), \quad \varphi^{\pr} = \varphi_l + {{2\pi }\o
{\b_0}} {{lN}\o {n-1}},
\label{4.43}
\er
where $s_a, \bar s_a$ and $L_a$ are integers and $q, \bar q, N=0, 1, \cdots n-2 \;\; {\rm mod }\;\; (n-1)$ are the $Z_{n-1}$
charges of the  corresponding fields.  Repeating the arguments of Sect. 4.2. we find that 
\br
w_-^{\pr} = w_- + {{\pi }\o {\b_0^2}} (s_- + {{N-2q}\o {n-1}}), \quad 
\bar w_-^{\pr} = \bar w_- + {{\pi }\o {\b_0^2}} (\bar s_- + {{N-2\bar q}\o {n-1}})
\label{4.44}
\er
and therefore the b.c.'s for the free field $\Phi_- = 2\b_0^2 {{n-1}\o {n+1}}w_-$ (i.e., $J_- = \pa \Phi_-$) are given by
\br
\Phi_- (\pm \infty ) = \pi (s_{\pm}^- + {{N_{\pm} -2 q_{\pm}}\o {n-1}})
\label{4.45}
\er
As a consequence the charges $Q_0^-$ and $\bar Q_0^-$ (\ref{4.19}) for the solutions having the above asymptotics take the
form, 
\br
Q_0^- &=& 4\pi {{n-1}\o {n+1}}(s_- + {{j_w^-}\o {n-1}}), \quad 
\bar Q_0^- = -4\pi {{n-1}\o {n+1}}(\bar s_- + {{j_{\bar w}^-}\o {n-1}}), \nonu \\
j_w^- &=& j_{\varphi} - 2 j_q, \quad j_{\varphi} = N_+-N_-, \quad s_- = s_+^- - s_-^-, \quad j_q = q_+ - q_-
\label{4.46}
\er
The explicit form $w_{top}^-$ of the $U(1)$-CFT topological solitons (carriyng $Q_0^-, \bar Q_0^-$  topological charges) is
again in the standard form (\ref{4.37}) with $\d_{-} = {{1}\o {\b_0^2}} (s_- + {{j_{\varphi} - 2 j_q}\o {n-1}})$.  Similarly
to the u-soliton case (\ref{4.402}), (\ref{4.41}) and (\ref{4.42}) but now according to the charges and energy relations
(\ref{4.21}), (\ref{4.24}) and (\ref{4.25}), we derive the following  semiclassical spectrum of the left moving 
1-soliton of the intermediate IM (\ref{2.16}):
\br
M_{int}^2 &=& M_g^{p=2} \( M_g^{p=2}  + \( {{16 m_0}\o {\b_0^2}} \) {{n-1}\o {n+1}} (s_- + {{j_{\varphi} - 2 j_q}\o {n-1}})^2\), \nonu
\\
Q_-^{el} &=& {{\b_0^2}\o {2}}j_{el}, \quad Q_+^{el} = \b_0^2 \d_0, \nonu \\
Q_{\bar R_u} &=& 4 {{n+1}\o {n-1}} \b_0^2 j_{el} + \pi (s_- + {{j_{\varphi} - 2 j_q}\o {n-1}}), \nonu \\
Q_{\theta}^{(1)} &=& - {{n+1}\o {8(n-1)}} Q_0^- = - Q^{(n)}_{\theta}
\label{4.47}
\er
where $Q_0^-$ is given by eqn. (\ref{4.46}).  The energy (and the mass $M_{int}(\a )$) for more general $\a$-dependent
$w_{top}^-(\a )$ (see eqn. (\ref{4.39})) are given by 
\br
E_{int}(\a ) &=& E_g^{p=2} + E_{L-sol}^{int}(\a ), \nonu \\
E_{L-sol}^{int}(\a ) &=& {{2|a_0|}\o {\b_0^2}}{{n-1}\o {n+1}}({{\pi }\o {\a}})^2 (1-\a \coth (\a ) ) (s_- + {{j_{\varphi} -2
j_q}\o {n-1}})^2, \nonu \\
M_{int}^2(\a ) &=& M_g^{p=2} \(  M_g^{p=2} + 2m_0 {{E_{L-sol}^{int} (\a )}\o {|a_0|}}\)
\label{4.48}
\er 
As in the case of the $U(1)$ IM (\ref{1.2}) (see Sect 3.5), the intermediate IM (\ref{2.16}) admits together with the
charged topological int-solitons of spectrum (\ref{4.47}) (or (\ref{4.48})) string-like solutions given by eqn. 
(\ref{4.23}), but with $w_{top}^- (\bar {w}_{top}^-)$ (ref{4.37})  replaced by
\br
w_{string}^- = w_{top}^- + w_{osc}
\nonu 
\er
They carry the same topological and electric charges (and new $Q_k^-, \bar Q_k^-, k\in Z$) (\ref{4.47}), but their energy
gets contributions from the oscillator part, i.e.,
\br
E_{int}^{string} = E_g + E_{L}^{string} + E_{R}^{string}
\nonu 
\er
The question of whether such classical string solutions are topologicaly stable (and remain stable under quantization)
is still open.  The fact that they have the same topological numbers but higher energies, is however an indication for their
topological instability.

\sect{Discussion and further developments}

Among the vast variety of integrable perturbations of the gauged $A_n$-WZW models we have chosen to study the soliton
solutions of a specific class of perturbations that preserve one or two chiral $U(1)$ symmetries and whose potentials have
n-distinct zeros.  Our main tool in the construction of the 1-solitons of these models is the nonconformal version of the
GKO coset construction \cite{god}, that allows us to compose these u-solitons in terms of massless solitons of the $U(1)$ (or 
$U(1)\otimes U(1)$)-CFT and already known g-solitons of the gauged $G_0/U(1)$  (or $G_0/U(1)\otimes U(1)$)
integrable models \cite{elek}, \cite{tau}, \cite{new}.  The new ingredients of this construction are the solitons and solitonic strings (left, right and
left-right )  of the corresponding CFTs presented in Sect. 3 and 4.  The simplest example ($n=1$) of $U(1)$ IMs (\ref{1.2})
written in the ``free field'' form
\br
{\cal L}_u^{n=1} &=& \pa \Phi \bar \pa \Phi + \b \bar \pa \g + \bar \b \pa \bar \g -\b \bar \b e^{-2\Phi} - 
m^2 \g \bar \g e^{2\Phi} + 2 \a_0 \Phi R^{(2)} \sqrt {-g}
\label{5.1}
\er
(where $R^{(2)}$ is the worldsheet curvature of the 2-d metric $g_{\mu \nu}, \; g = det g_{\mu \nu}$), is known to represent
an integrable perturbation of string on $AdS_3$ target space.
One expects that $ AdS_3/ CFT_2$ Maldacena correspondence \cite{mal} takes place for the relevant deformed theories, i.e. the 
$AdS_3$ with perturbation $m^2 \g \bar \g e^{2\Phi}$ to be equivalent to the deformed $CFT_2$ (i.e. 2-d integrable model)
{living } on the border of the $AdS_3$ space.  
Since the energy $\tilde E$ and angular momenta $\tilde l$ of the latter theory are related to the charge $Q_0$ and $\bar
Q_0$ of the $AdS_3$ model (\ref{5.1}) \cite{mo}, \cite{b}, 
\br
Q_0 = {1\o 2} (\tilde E + \tilde l), \quad \quad \bar Q_0 = {1\o 2} (\tilde E - \tilde l)
\label{5.2}
\er
the charge spectrum of (\ref{5.1}) we have derived in Sect. 3 determines  the energy spectrum of the border deformed $CFT_2$. 
One can consider the $n\geq 2$ IMs (\ref{1.2}) as specific integrable deformations of the bosonic string on 
$AdS_3 \otimes T_{n-1}$ target space which provides its charge spectrum (\ref{3.36}) and (\ref{3.52}) with certain string
(and border $CFT_2$) meaning.  More interesting examples are given by the IMs (\ref{2.13}) studied in Sect. 4.
 They can be considered as 
integrable relevant perturbations of the $AdS_3 \otimes S_3 \otimes T_{n-1}$ string model (taking $R_n \rightarrow iR_n$ and
$\psi_n^{*} = \chi_n$).    It is important  to mention that the spectral flow (\ref{3.58}) of charges $Q_0, \bar Q_0$ is
realized  by adding topological $\theta$-terms (\ref{3.54}) to the original deformed string Lagrangians (\ref{1.2}).

As it well known the standard procedure of restoring the conformal invariance of the deformed IM (\ref{5.1}) (and
(\ref{1.2}) in general) is to consider their conformal affine versions \cite{2loop}, \cite{bon}. 
 By introducing a new pair of fields ($\nu ,
\eta $) one can map the nonconformal IM (\ref{5.1}) to the following conformal integrable model
\br 
{\cal L}_{CAT}^{n=1} &=& \pa \Phi \bar \pa \Phi + \b \bar \pa \g + \bar \b \pa \bar \g + \pa \nu \bar \pa \eta + \pa \eta
\bar \pa \nu \nonu \\
&-& \b \bar \b e^{-2\Phi } -m^2 \g \bar \g e^{2\Phi - \eta } + 2 ( \a_0 \Phi + \g_0 \nu )R^{(2)} \sqrt {-g}
\label{5.3}
\er
where $\a_0$ and  $\g_0$ are static background charges.
An important feature of such IM is that the free field $\eta$  
plays  the role of renormalization group parameter that interpolate between different conformal backgrounds (i.e. the zeros
of the corresponding $\s $-model $\b$ -functions):
\begin{itemize}
\item 
$\eta \rightarrow \infty $ leads back to the original conformal $SL(2,R)$-WZW model 
\item
$\eta \rightarrow 0 $ is reproducing the relevant perturbation  (\ref{5.1})
\end{itemize}
The model (\ref{5.3}) admits new massless solitons characterized now by the  topological charge 
\br
Q_{\eta} = \b_0 \int_{-\infty}^{\infty} \pa_x \eta dx
\nonu 
\er
It is interesting to derive their complete energy and charge spectrum as well as to understand the string meaning of the CAT
versions of the solitonic string solutions of (\ref{5.1}).

An interesting example of an off critical bosonic string on curved target space of black hole 
type \cite{a} is given by the
$n=2$ intermediate IM (\ref{2.16}) (with local $U(1)$ symmetry):
\br
{\cal L}^{int}_{n=2} &=&{1\o 3} \pa \bar R \bar \pa \bar R + 
{{\sum_{a,b=1}^{2} f_{ab} \( (1+ {{\b^2}\o 4} \bar \psi_a \bar \chi_a
e^{\b \eps_a \bar R} ) \pa \bar \chi_b \bar \pa \bar \psi_b e^{\b \eps_b \bar R} 
- \bar \psi_a \bar \chi_b \bar \pa \bar \psi_b \pa \bar
\chi_a\) } \o {1+{{\b^2}\o 4}  (\bar \psi_1 \bar \chi_1 e^{\b  \bar R}+ \bar \psi_2 \bar \chi_2 e^{-\b  \bar R})}} \nonu \\
&-& m^2 (\bar \psi_1 \bar \chi_1 e^{\b  \bar R}+ \bar \psi_2 \bar \chi_2 e^{-\b  \bar R} + 
\b^2 \bar \psi_1 \bar \chi_1\bar \psi_2 \bar \chi_2)
\label{5.32}
\er
where $f_{aa} =0, \; f_{12} = f_{21} = 1, \;\; \eps_1 = -\eps_2 = 1$.  For $m^2 =0$, the 
$\s $-model (\ref{5.32}) is
conformal invariant.  Since corresponding target space metric  admits both horizons and 
singularities, it represents a
specific 5-d generalization of the 3-d black string model \cite{horne}.  For $m^2 \neq 0$ the
 above nonconformal IM has
$U(1)$-charged massive and massless solitons (and strings), constructed in Sect.4.3, that can 
be interpreted as charged 
strong coupling (stable) particles (and strings).  Its CAT version (i.e. with $\nu$ and $\eta $ 
as in eqns. (\ref{5.3}))
 might  play an important role  in the understanding of the nonperturbative properties of the  
 IM (\ref{5.32}) (and of the
 corresponding string models ) due to the fact that  
 the massless  solitons introduces new nonperturbative string states.

Another direction of extending the results of the present paper is to 
consider IMs with {\it nonabelian} local symmetries.  The
simplest example of integrable perturbation of the $SL(3,R)$-WZW model 
that preserve chiral $SL(2,R)\otimes U(1)$ symmetry
was derived in our recent paper \cite{new}
\br
{\cal L}_{SL(3,R)\;\; pert} &=& {\cal L}_{WZW}^{SL(3,R)}(g_0) - 
{{m^2}\o {\b^2}}\( {2\o 3} + \psi_1 \chi_1e^{\b (2\varphi_2
- \varphi_1)} + \psi_2 \chi_2 e^{\b (\varphi_1+\varphi_2)}\) 
\label{5.4}
\er
 where the $SL(3,R)$ group element is parametrized  as follows:
\br
g_0 = e^{\chi_3 E_{-\a_1}}e^{\chi_1E_{-\a_2}+\chi_2E_{-\a_1-\a_2}}e^{\varphi_1 h_1 + \varphi_2 h_2} 
e^{\psi_1E_{\a_2}+\psi_2E_{\a_1+\a_2}}e^{\psi_3 E_{\a_1}}
\nonu
\er
 By the methods of Sect. 2 one can construct integrable perturbations of $SL(n+1)$-WZW model keeping unbroken certain
 subgroup $G_0^0 = SU(k), k\leq n$.  We expect that the nonconformal GKO coset construction of Sect. 3 (for the particular
 cases $G_0^0 = U(1)$ or $G_0^0 = U(1)\otimes U(1)$) to take place for arbitrary nonabelian $G_0^0$ as well.  As in the
 $U(1)$ case, the $G_0^0$-CFT massless solitons should be the main ingredient in the construction of the 1-solitons of such
 IM, as (\ref{5.4}) for example.  The derivation of their energy and charge spectrum (say for $G_0^0 = SL(2)$) is an
 interesting open problem.
 
 Our last comment concerns the nonrelativistic analogs of IM (\ref{1.2}).  The question is 
 whether one can construct {\it nonrelativistic} IMs
 with local $U(1)$ symmetry.  The well known relation between Lund-Regge (L-R) IM ($n=1$ of eqn. (\ref{1.9})) and the nonlinear
 Schroedinger model \cite{aratyn1} and the fact that L-R model is the gauged version of the deformed $SL(2,R)$ WZW (\ref{5.1})
address the question about  the nonrelativistic counterpart (in the sense of ref. \cite{wigner}, \cite{new}) of the IM (\ref{5.1}).  The
answer to this question is given by the following integrable system of second order differential equations for the field
variables $r(x,t), q(x,t) $ and $u(x,t)$:
\br
\pa_t r - {1\o 2}\pa_x^2 r + 2 u \pa_x r + r \pa_x u -2r (u^2 -rq)&=&0, \nonu \\
\pa_t q + {1\o 2}\pa_x^2 q + 2 u \pa_x q + q \pa_x u +2q (u^2 -rq)&=&0,\nonu \\
\pa_t u + {1\o 2} \pa_x (rq ) &=&0
\label{5.5}
\er
which is invariant under local $U(1)$ transformations:
\br
u=u^{\pr} + {{\b }\o 2}\pa_x w(x), \quad \quad r= e^{\b w(x)}r^{\pr}, \quad \quad q= e^{-\b w(x)}q^{\pr}
\label{5.6}
\er
We expect that its solitons are related  to the Non linear Schroedinger (NLS) solitons in the same way the u-solitons of the IM (\ref{5.1})
are related to the L-R nontopological solitons.  
The generalization of the IM (\ref{5.5}) to the family of nonrelativistic IMs with local $U(1)$
symmetry related to the relativistic IMs (\ref{1.2}) is straightforward.

{\bf  Acknowledgements}
We are grateful to R. Paunov for discussions concerning integrable perturbations of SL(2,R) WZW model.We thank
 Fapesp, CNPq and Unesp for partial financial support.

\end{document}